\documentclass[12pt,usenames,dvipsnames,a4paper,final]{article}

\usepackage{units}

\usepackage[nohead,includefoot,includemp,textwidth=200mm,hoffset=-50pt,marginparsep=2pt,marginparwidth=95pt] {geometry}
\usepackage[font=footnotesize,labelfont=bf]{caption}

\newcommand{\blue}{\color{blue}}

\newcommand{\dat}{{\rm dat}}

\renewcommand{\Re}{\mathop{\mathrm{Re}}}
\renewcommand{\Im}{\mathop{\mathrm{Im}}}
\newcommand{\ket}[1]{|#1\rangle}
\newcommand{\bra}[1]{\langle #1|}
\usepackage{datetime} 
\usepackage{subfig}
\usepackage{makeidx}
\usepackage{graphicx}
\usepackage{paralist}

\def\Xint#1{\mathchoice
   {\XXint\displaystyle\textstyle{#1}}%
   {\XXint\textstyle\scriptstyle{#1}}%
   {\XXint\scriptstyle\scriptscriptstyle{#1}}%
   {\XXint\scriptscriptstyle\scriptscriptstyle{#1}}%
   \!\int}
\def\XXint#1#2#3{{\setbox0=\hbox{$#1{#2#3}{\int}$}
     \vcenter{\hbox{$#2#3$}}\kern-.5\wd0}}

\def\dashint{\Xint-}

\usepackage{amsmath,bbold,amsfonts,stmaryrd,bm}

\usepackage{appendix}
\usepackage{amssymb}
\usepackage{float}
\usepackage{color} 
\usepackage{showlabels}

\usepackage[draft,textsize=scriptsize,color=ForestGreen!20,prependcaption]{todonotes}%

\usepackage{hyperref}
\newcommand\myshade{85}
\colorlet{mylinkcolor}{violet}
\colorlet{mycitecolor}{Aquamarine}
\colorlet{myurlcolor}{YellowOrange}

\hypersetup{
 linkcolor  = mylinkcolor,
 citecolor  = mycitecolor!\myshade!black,
 urlcolor   = myurlcolor!\myshade!black,
colorlinks = true,
}

\def\a{\alpha}

\def\b{\beta}

\def\d{\delta}
\def\D{\Delta}

\def\eps{\varepsilon}
\def\f{\frac}

\def\G{\Gamma}

\def\l{\left}

\def\la{\langle}
\def\ra{\rangle}

\def\mc{\mathcal}

\def\m{\mu}
\def\n{\nu}

\def\p{\partial}

\def\r{\right}
\def\s{\sigma}
\def\t{\theta}

\def\Tr{\mathrm{Tr}}

\def\dst{\displaystyle}

\def\be{\begin{equation}}
\def\ee{\end{equation}}

\def\bea{\begin{eqnarray}}
\def\eea{\end{eqnarray}}

\def\ba{\begin{array}}
\def\ea{\end{array}}

\def\bc{\begin{center}}
\def\ec{\end{center}}

\def\bl{\begin{flushleft}}
\def\el{\end{flushleft}}

\def\br{\begin{flushright}}
\def\er{\end{flushright}}

\def\bi{\begin{itemize}}
\def\ei{\end{itemize}}

\def\bt{\begin{tabular}}
\def\et{\end{tabular}}

\begin{document}

\thispagestyle{empty}

\title{Hadronic decays of a light Higgs-like scalar}

\author{Alexander~Monin$^1$\thanks{\texttt{\small alexander.monin@unige.ch}},
Alexey~Boyarsky$^2$,
Oleg Ruchayskiy$^3$\\
$^1${\small Department of Theoretical Physics, University of Geneva} \\
{\small 24 quai Ernest-Ansermet, 1211 Geneva} \\
$^2${\small Lorentz Institute, Leiden University,}\\
{\small Niels Bohrweg 2,
Leiden, NL-2333 CA, The Netherlands}\\
$^3${\small Discovery Center, Niels Bohr Institute, Copenhagen University,}\\
{\small Blegdamsvej 17, DK-2100 Copenhagen, Denmark}}
\date{}
\maketitle
\begin{abstract}
We revisit the question of hadronic decays of a GeV-mass Higgs-like scalar.
 A number of extensions of the Standard Model predict Higgs sector with additional light scalars.
 Currently operating and planned Intensity Frontier experiments will probe for the existence of such particles, while theoretical computations are plagued by uncertainties.
 The goal of this paper is to bring the results in a consolidated form that can be readily used by experimental groups.
 To this end we provide a physically motivated fitting ansatz for the decay width that reproduces the previous non-perturbative numerical analysis. We describe systematic uncertainties of the non-perturbative method and provide explicit examples of the influence of extra resonances above $1.4$~GeV onto the total decay width.
\end{abstract}

\section{Motivation}

The Standard Model of particle physics provides a closed and self-consistent description of known elementary particles interacting via strong, weak and electromagnetic forces.
The Standard Model coupled with general relativity was also very successful in describing the evolution of the Universe as a whole.
However, the impressive success of the Standard Model at accelerator and cosmic frontiers also revealed with certainty that the Standard Model fails to explain a number of observed phenomena in particle physics, astrophysics and cosmology.
These major unsolved challenges are commonly known as ``beyond the Standard Model'' (BSM) problems.
They include: neutrino masses and oscillations, dark matter, baryon asymmetry of the universe, etc.

A range of possible scenarios capable to resolve the BSM puzzles is extremely wide. 
At the one end there are models such as $\nu$MSM (Neutrino Minimal Standard Model)~\cite{Asaka:2005an,Asaka:2005pn} that postulate only 3 extra particles lighter than electroweak scale, providing resolutions of major BSM puzzles and leading to a Standard Model-like quantum field theory up to very high scales~\cite{Shaposhnikov:2007nj,Bezrukov:2012sa,Bezrukov:2014ina}.
At the other end there are models where one is completely agnostic about the
structure of the hidden (``dark'') sectors and explores \emph{portals} -- mediator particles, that both couple to states in the ``hidden sectors'' and interact with the Standard Model.
Such portals can be renormalizable (mass dimension $\le 4$) or be realised as higher-dimensional operators suppressed by the dimensionful couplings $\Lambda^{-n}$, with $\Lambda$ being the new energy scale of the hidden sector.
Mediator couplings to the Standard Model sector can be sufficiently small to allow for the portal particles to be (much) lighter than the electroweak scale.
Such models can be explored with \emph{Intensity} (rather than \emph{Energy}) frontier experiments.

In this paper we focus on scalar (or ``Higgs'') portal~\cite{Patt:2006fw} -- gauge singlet scalar $S$ interacting with the Higgs doublet $H$ via $S H^\dagger H$ term.
Such particles ``inherit'' their interactions from the Higgs boson (albeit suppressed by a small dimensionless parameter $\theta$).
New generation of Intensity Frontier experiments, such as NA62~\cite{Flacke:2016szy,Mermod:2017ceo,Dobrich:2017yoq}, SHiP~\cite{Alekhin:2015byh,Anelli:2015pba}, MATHUSLA~\cite{Curtin:2017izq,Evans:2017lvd}, FASER~\cite{Feng:2017vli}), CODEX-b~\cite{Gligorov:2017nwh}, SeaQuest~\cite{Berlin:2018pwi} will probe for existence of such scalars with the masses $\sim\text{GeV}$.
The lifetime of such scalars is dominated by the decay into light mesons ($S\to \pi\pi$, $S\to \bar K K$, etc.)
The question of computation of the decay width of such particles has been studied in '80s~\cite{Anselm:1985wg,Voloshin:1985tc,Raby:1988qf,Chivukula:1989ds,Chivukula:1989ze,Truong:1989my, Donoghue:1990xh} in the context of hadronic decays of light Higgs boson.
Based on the data for $\psi'\to \psi \pi \pi$ and $\Upsilon'\to \Upsilon \pi \pi$ decays, \cite{Voloshin:1985tc} argued in favor of extrapolating the results obtained with the help of Chiral Perturbation Theory (ChPT) up to $1.5$~GeV. At the same time the non-perturbative analysis of~\cite{Donoghue:1990xh} produced results differing from~\cite{Voloshin:1985tc}
by as much as an order of magnitude.

This discrepancy, crucial for the new generation of experiments, warrants the current work.
We critically review existing methods of computation of the scalar's hadronic width and assess the uncertainties.
We mainly reconfirm findings of~\cite{Donoghue:1990xh} but provide a way to
assess its uncertainties and speculate up to what scales the non-perturbative
approach should be used {(read trusted).}

The paper is organized as follows.
In Section~\ref{sec:setup} we discuss briefly the properties of scalar portal and define the form-factors through which the hadronic decay width $\Gamma_{\pi\pi}$ is expressed.
We review computation of the hadronic decay in the chiral perturbation theory, reproducing the results of~\cite{Voloshin:1985tc} in Section~\ref{sec:ChPT}.
The unitarity arguments that allow for non-perturbative treatment of the relevant form-factors are summarized in Section~\ref{sec:beyond-ChPT}.
The review of dispersive methods is given in Section~\ref{sec:review-methods}.
Section~\ref{sec:results} summarises our results and compares with previous works.
Section~\ref{sec:discussion} provides error estimate and domain of validity.
We conclude in Section~\ref{sec_conclusion} and provide supplementary material in Appendices.

\section{Setup \label{sec:setup}}

We start from laying down the ground rules for computing the desired decay rate. We consider a scalar field $S$ weakly coupled to the Standard Model Higgs field $H$, see~\cite{Alekhin:2015byh} for details. For masses below $1$~GeV the relevant UV couplings of the scalar are only those to quarks and leptons
\begin{equation}
\mc L _ {int} = - \f {S} {v_S} \sum _ {q } m _ q \bar q q - \f {S} {v_S} \sum _ {\ell } m _ \ell \bar \ell \ell,
\label{eq:scalarUVInteractions}
\end{equation}
since the only interesting decay channels are $\pi\pi$, $\mu^+ \mu^-$ and possibly $\bar K K$.
In Eq.~\eqref{eq:scalarUVInteractions} $v_S \equiv v \cot\theta$, where $v = 246$~GeV is the Higgs vev and $\theta \ll 1$ parametrises the interaction of the scalar $S$ with the Standard Model particles.

From the Lagrangian~\eqref{eq:scalarUVInteractions}  the decay rate $S \to \mu^+ \mu^-$ can be immediately found
\begin{equation}
\G _{\mu^+ \mu^-} = \f {1} {8 \pi} \f {m_\m^2 m_S} {v_S^2} \l ( 1 - \f {4 m_\m^2} {m_S^2}\r ) ^ {3/2}.
\label{eq:muonDecay}
\end{equation}
Computing the width due to hadronic decays is somewhat more involved. The difficulty stems from the strong coupling of QCD in the regime of interest. Therefore, considering only the tree level process -- which in the case of the leptonic decay leads to (\ref{eq:muonDecay}) -- is not enough. Quarks and gluons are not adequate degrees of freedom for describing the low energy physics. Instead, in order to compute hadronic decay rates, matrix elements of the Lagrangian (\ref{eq:scalarUVInteractions}) between low energy hadronic states should be computed directly. For instance, in the case of $S\to \pi^a \pi^b$, where $a$ and $b$ are isospin indices, the amplitude is defined as
\begin{equation}
\mc A _{\pi} (m_S^2) \d^{ab} \equiv \la S | \mc L _ {int} | \pi^a (p _{ 1}) \pi^b ({p_{2}}) \ra = - \f {1} {v _ S} \la 0 | \sum_q m_q \bar q q | \pi^a (p _{ 1}) \pi^b ({p_{2}}) \ra.
\label{eq:pipiMatrixElement}
\end{equation}
Integrating it over the phase space gives  the following decay width
\begin{equation}
\G _{\pi\pi} = \f {3} {32\pi} \f{\bigl | \mathcal A _{\pi} (m_S^2)\bigr | ^2} {m_S} \sqrt{1 - \f {4 m_\pi ^2} {m_S^2}},
\label{decay_rate_pipi}
\end{equation}
where we summed over all species and took into account that the particles in the final state are identical.

The sum in (\ref{eq:pipiMatrixElement}) contains contributions from light ($u$, $d$, $s$) and heavy ($c$, $b$, $t$) quarks. The latter can be expressed in terms of the former and the energy momentum tensor by using a clever trick based on the knowledge of the trace anomaly and the RG invariance of the energy momentum tensor (for more details see~\cite{Shifman:1978zn,Vainshtein:1980ea,Voloshin:1985tc} and \cite{Chivukula:1989ds,Chivukula:1989ze,Kaplan:2000hh}). It uses two different representations of the energy momentum tensor at energies immediately above and below the $c$-quark mass. On the one hand using the UV description (all quarks) the trace of the energy momentum tensor, due to the anomaly, is given by
\begin{equation}
\theta ^{ \m }_{\m} = \f {\beta (\a_s)} {4 \a_s} G ^{ 2} + \sum _ {\text{all}} m _ q\bar qq, ~~ \a_s = \f {g_s^{2}} {4 \pi}, ~~ G ^{ 2} = G ^{ a} _{\m \n} G ^{ a} _{\m \n},
\label{conf-an}
\end{equation}
with the one-loop beta function for the strong coupling $\a_s$ defined as
\begin{equation}
\b ( \a_s) = -\f {b\a_s ^ 2 }{2 \pi}, ~~ b = 9 - \f {2} {3} N _{ h},
\end{equation}
where $N_{h}$ being the number of heavy quarks (in our case 3). On the other hand from the IR perspective (after integrating out heavy quarks) the energy momentum tensor becomes
\begin{equation}
\t _{\m}^{\m}= \f {\bar \beta (\a_s)} {4 \a_s} G ^{ 2} + \sum _ {\text{light}} m _ q \bar qq + O (1/m_c^2),
\end{equation}
where the reduced beta function corresponds to only light quarks (u,d,s)
\begin{equation}
\bar \b ( \a_s) = -\f {9\a_s ^ 2 }{2 \pi}.
\end{equation}
As a result one concludes that
\begin{equation}
 \sum _ {\text{heavy}} m _ q \bar qq = - \f {2}{3}N _{ h} \f {\a_s}{8 \pi} G ^{ 2} = \f {2} {27} N _{ h} 
 \l (\theta_{\m}^{\m} - \sum _ {\text{light}} m _ q \bar qq \r ),
\label{FF-G2}
\end{equation}
and the interaction Lagrangian can therefore be rewritten as
\begin{equation}
\mc L _{int} = - \f {S} {v_S} \l [ \f {2} {27} N _{h} \t _{\m}^{\m}+ \l ( 1 -  \f {2} {27} N _{h} \r )\sum _ {\text{light}} m _ q \bar qq \r ],
\label{h-qq-Ch}
\end{equation}
which immediately leads to the following expression for the amplitude (\ref{eq:pipiMatrixElement}) in the leading order in $\a_s$
\begin{equation}
\mc A _{\pi} (m_S^2)= \f {i} {v _ S} 
\l \{ \f {2} {27} N _{h} \theta_\pi (m_S^2) + \l [ 1 -  \f {2} {27} N _{h} \r ] \l [ \Gamma_\pi (m_S^2) + \D_\pi (m_S^2) \r ] \r \},
\label{eq:SpipiAmplitude}
\end{equation}
where the following notations for the form factors were introduced
\begin{subequations}
\begin{align}
\label{ff_Gamma}
\Gamma_\pi(s) \d^{ab} &= \la \pi^a \pi^b | m _{ u} \bar u u + m _{d} \bar d d | 0 \ra, \\
\label{ff_Delta}
\D_\pi(s) \d^{ab} &= \la \pi^a \pi^b | m _{ s} \bar s s | 0 \ra, \\
\label{ff_theta}
\t _\pi(s) \d^{ab} &= \la \pi^a \pi^b | \theta_{\m}^{\m} | 0 \ra.
\end{align}
\end{subequations}
The problem of computing the width (\ref{decay_rate_pipi}) thus boils down to computing these form factors. In the next sections we present several approximations when it can be done using different techniques such as the ChPT and unitarity. It is also important to note that the expression (\ref{eq:SpipiAmplitude}) does not capture effects suppressed by heavy quark masses $1/m_c^{2}$.

\section{ChPT}
\label{sec:ChPT}

For very small energies the form factors (\ref{ff_Gamma}-\ref{ff_theta}) can be easily computed using the chiral perturbation theory~\cite{Voloshin:1985tc,Voloshin:1980zf}. We will not go to great lengths to introduce the ChPT referring the reader to numerous sources (for instance~\cite{Weinberg:1996kr,Scherer:2002tk,Scherer:2012xha}). Instead we give just the key results allowing to demonstrate how the computation is done.

The low energy dynamics of QCD can be described (in the case of the $SU(2)\times SU(2)$ chiral symmetry) by introducing an $SU(2)$ matrix
\begin{equation}
\Sigma = e^ {i \s_a \pi_a/f_\pi},
\end{equation}
parametrized by pion fields $\pi_a$, with $f_\pi=93$ MeV being the pion decay constant. The action of the chiral symmetry group on the space of these matrices is realized by the right and left multiplications
\begin{equation}
\Sigma' = U_L \Sigma U^\dagger _R.
\label{su2_chirla_sym}
\end{equation}
In building the Lagrangian one has to make sure that the symmetry (\ref{su2_chirla_sym}) is preserved. It is straightforward to show that the leading (derivative expansion) order Lagrangian is given by
\begin{equation}
\mc L = \f {f_\pi^2} {4} \Tr \p _ \m \Sigma \p_\m \Sigma^\dagger + \f {B f_\pi ^ 2} {2} \Tr ( M^ \dagger \Sigma + \Sigma ^ \dagger M),
\end{equation}
where $M$ is the quark mass matrix
\begin{equation}
M = \l ( 
\begin{array}{ccc}
m_u & 0 \\
0 & m _ d
\end{array}
\r ),
\end{equation}
and $B$ is a constant. Expanding the Lagrangian up to quadratic order shows that the pion mass is given by $m_\pi^2 = B (m_u + m_d)$.
At this order the trace of the energy momentum tensor is given by
\begin{equation}
\theta_\m^\m = 2 m_{\pi}^2 \pi^2 - (\p\pi)^2.
\end{equation}
Therefore, the corresponding form factor (\ref{ff_theta}) becomes
\begin{equation}
\theta_\pi(s) = s+2m _\pi^2.
\label{low-energy-theta-ff}
\end{equation}
The other two form factors (\ref{ff_Gamma}) and (\ref{ff_Delta}) can be computed in the similar way
\begin{equation}
\Gamma_\pi(s) = m_{\pi}^{2}, ~~ \D_\pi(s) = 0.
\label{low-energy-ff}
\end{equation}
As a result the amplitude (\ref{eq:SpipiAmplitude}) becomes~\cite{Voloshin:1985tc}
\begin{equation}
\mc A _\pi = i \f {2} {9} v_S \l ( m _{S}^{2} + \f {11} {2} m _{\pi}^{2} \r ),
\label{LO-apm}
\end{equation}
which -- upon using (\ref{decay_rate_pipi}) -- leads to the following expression for $S \to \pi \pi$ decay rate 
\begin{equation}
\G _ {\pi \pi} = \f {1} {32\pi} \sqrt{1 - \f {4 m_\pi^2} {m_S^2}} \f {4} {81 m_S v_S^2} \l ( m_S^2 + \f {11} {2} m _ \pi ^ 2 \r )^2.
\end{equation}

By construction the ChPT is only reliable for sufficiently small energies. It is thus clear that the result (\ref{LO-apm}) is valid at energies low compared to QCD scale, i.e. $\Lambda_{\text{QCD}}\equiv4\pi f _\pi \approx 1$ GeV. At the same time we are interested in finding the decay rate for scalars with masses comparable to 1 GeV. In order to do that it is not enough to compute the next to leading order correction in ChPT. In the next section we'll present a non-perturbative approach based on dispersion relations.

\section{Beyond ChPT and unitarity}
\label{sec:beyond-ChPT}

In the previous section we showed how the decay rate of a light scalar into pions can be computed using the power of the effective field theory. ChPT corrections to the leading order result are suppressed by powers of $s/\Lambda_{\text{QCD}}^{2}$. Therefore, perturbative computations become unreliable at energies close to the cutoff. However, the precise point where corrections become comparable with the leading order computation depends on the specifics of an observable. There are indications that for the form factor $\Gamma_\pi(s)$ it happens for energies $s$ much smaller\footnote{It is known that final state interaction effects can be rather strong~\cite{Roiesnel:1980gd}.} than $\Lambda_{\text{QCD}}^2$.

Using the definition of the quadratic scalar radius of the pion
\begin{equation}
\la r^{2} \ra_{S,\pi} = 6\f {\p \log \Gamma_\pi(s)} {\p s} \Big |_{s=0},
\end{equation}
the form factor $\G_\pi(s)$ around $s=0$ can be written as
\begin{equation}
\Gamma_\pi (s) = \Gamma_\pi (0) \l ( 1 + \f {1} {6} s \la r^{2} \ra_{S,\pi} + \dots \r ),
\label{scalar_radius}
\end{equation}
where $ \Gamma_\pi (0)$ is given by Eq.~\eqref{low-energy-ff}. The quadratic scalar radius of the pion was first computed in~\cite{Gasser:1984ux} using ChPT at one loop, with the result $\la r^{2} \ra_{S,\pi} = 0.55 \pm 0.15 \text{ fm}^{2}$. The method of~\cite{Donoghue:1990xh} (to be discussed shortly) produced $\la r^{2} \ra_{S,\pi}=0.600 \pm 0.052 \text{ fm}^{2}$. Later using a better input for the pi-pi phases the analysis was repeated in~\cite{Colangelo:2001df}, producing
\be
\la r^{2} \ra_{S,\pi}=0.61 \pm 0.04 \text{ fm}^{2}.
\label{eq:qsrLattice}
\ee
Recently this computation was corroborated by lattice computations in~\cite{Gulpers:2015bba}\footnote{It is precisely the result of~\cite{Colangelo:2001df} and lattice computations that will be used in the following sections to fix unknown coefficients in several form factors.}. Such a value of the quadratic scalar radius implies that ChPT cannot be trusted when $\f {1} {6} s \la r^{2} \ra_{S,\pi} \sim 1$ or, equivalently, at $\sqrt{s} \sim$ 600--700 MeV (see also~\cite{Chivukula:1989ds,Chivukula:1989ze}). Therefore, for masses of a scalar $m_S \lesssim 1$~GeV a non-perturbative approach should be used.

\subsection{Analyticity and unitarity}
\label{sec:unitarity}

Such a method by definition should use only the most general constraints on form factors without alluding -- if possible -- to any specific perturbative computation. The first constraint comes from analyticity. It can be proven (see books~\cite{Bjorken:1965zz,Barton:1965}) that form factors are analytic functions in the complex plane of the variable $s$ with the cut $s>4m_\pi^2$, and it can be established (using high energy behavior of QCD~\cite{Yndurain:2002ud}) that their behavior at infinity is $\phi_i(s) \sim 1 / {s}$.

The second constraint is due to unitarity. To discuss it we have to introduce the notion of the scattering matrix for s-waves with isospin zero. For two-to-two 
($\pi^a\pi^b\to\pi^c\pi^d$) scattering, the $S$-matrix defined as,
\begin{equation}
S_{abcd}(s,t,u) = {}_{\text{out}}\langle \pi_c (p_3) \pi_d (p_4)  | \pi_a (p_1) \pi_b (p_2) \rangle_{\text{in}}
\label{S-matrix_gen}
\end{equation}
depends on all Mandelstam variables ($s$, $t$ and $u$) and has an arbitrary tensor structure in the space of isospin indices $a,\dots, d$. However, it can be expanded in partial waves with fixed angular momentum $J$ and isospin $I$. We are interested in scalar (isoscalar) form factors, therefore, we consider only s-wave isospin zero ($J=I=0$) scattering, by projecting (\ref{S-matrix_gen}) on the corresponding subspace (see Chapter 19 in~\cite{Weinberg:1996kr}). It is this component that we refer to as $S$-matrix in what follows. For energies below the inelastic threshold ($4m_\pi$), the $S$-matrix is completely determined by the pion phase shift
\begin{equation}
S (s) =e ^{2i \d _{\pi} (s)}, ~~ 4 m_{\pi}^{2}<s<16 m_{\pi}^{2}.
\label{1ch_S-matrix}
\end{equation}
As energy grows, channels $2\pi \to 4\pi$, $2\pi \to 6\pi$ and then $2\pi \to \bar K K$ open up.
Correspondingly, the $S$-matrix can be represented by a finite dimensional matrix, $S_{ij}$, with $i$ and $j$ running in the space of channels~\cite{pdg_resonances}. It is observed experimentally that the mixing with multi-particle (four and more) states for energies below $\Lambda_{\dat} = 1.4-1.6$~GeV is small~\cite{Hyams:1973zf,Au:1986vs}.
Therefore, in this region there are effectively only two relevant channels $(\pi\pi\to\pi \pi, \pi\pi\to\bar K K)$.

\begin{figure}[h]
 \centering \subfloat[][Schematic representation for the
  unitarity condition: S-matrix]{\label{fig:2chSMatrix}\includegraphics[height=3cm]{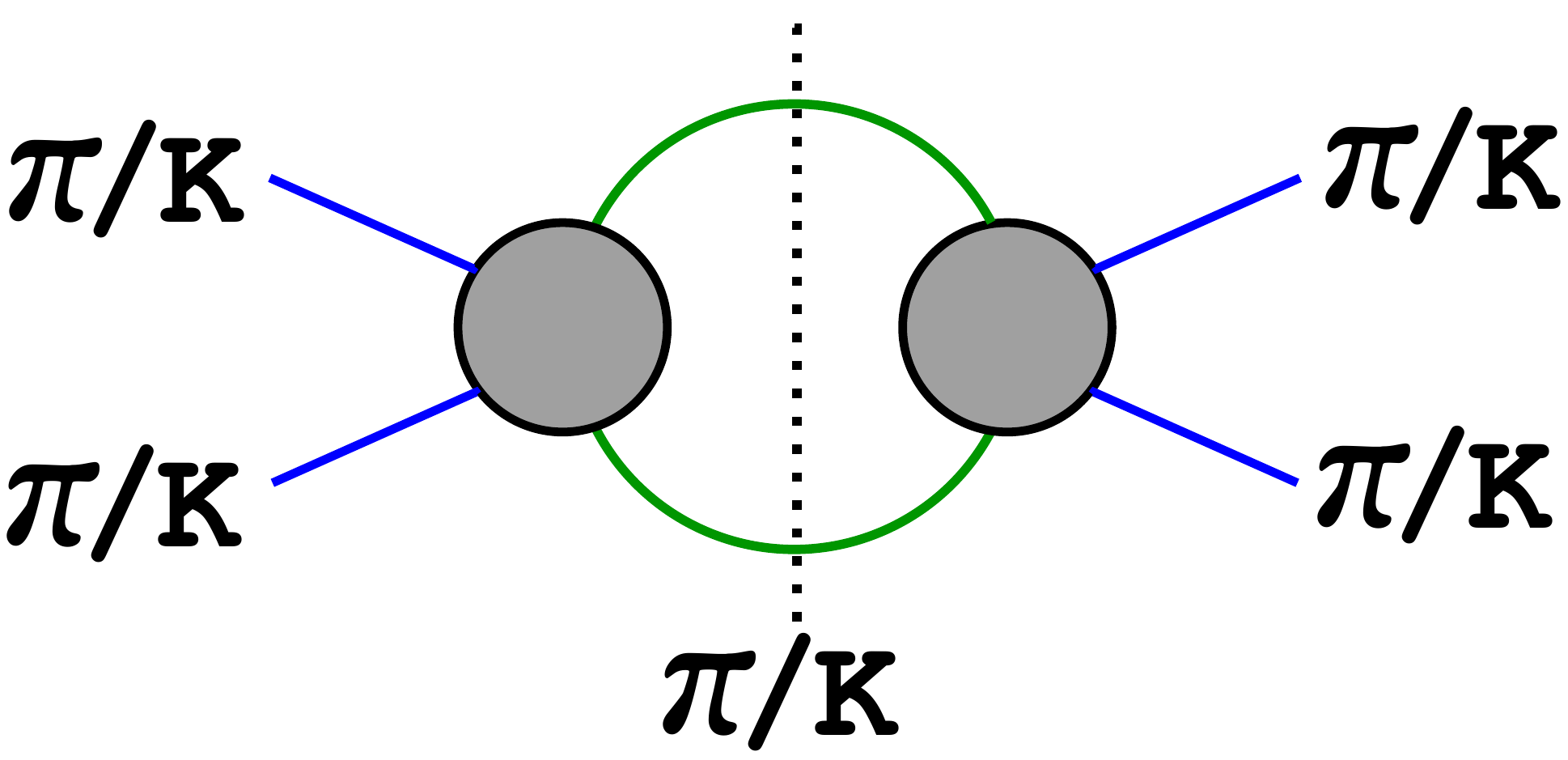}}~~~~%
\subfloat [][Specialization of unitarity condition for the case of form-factors $\phi_i(s)$ in the two-channel approximation]{\label{fig:2chFormFactors} \includegraphics[height=3cm]{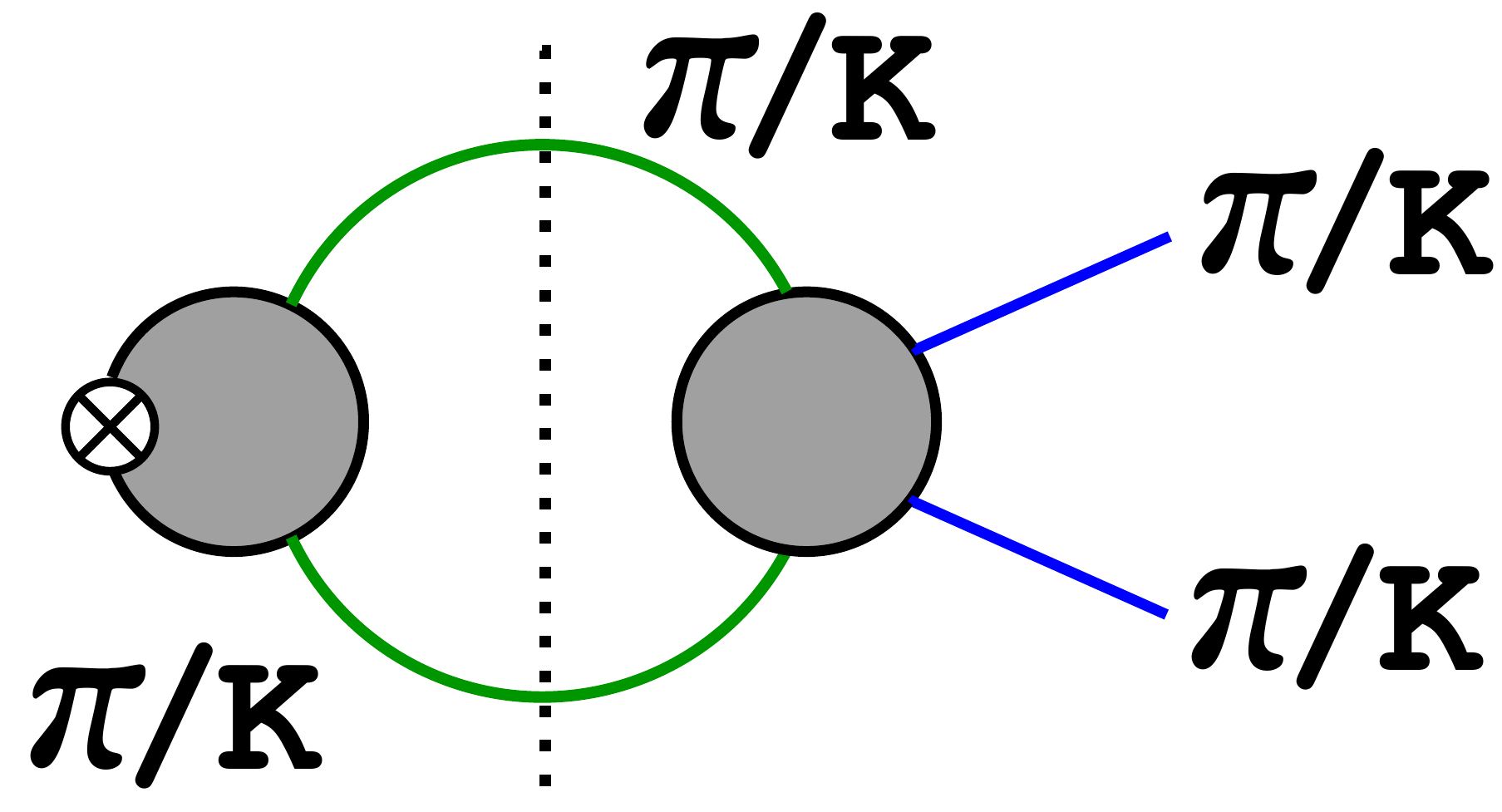}}
  \caption{\label{cut-general}Unitarity conditions for form factors and $S$-matrix}
\end{figure}

In the case of two channels -- the generalization to an arbitrary number of channels seems straightforward -- unitarity constraints for form factors, similar to the optical theorem (see e.g.\ the textbook~\cite[Sec.~6-3-4]{Itzykson:1980rh}) for amplitudes, can be derived in the following way. We define 
\begin{equation}
  \begin{aligned}
    \phi_1(s) \d^{ab}&\equiv \bra{\pi^a\pi^b} X \ket{0}, \\
    \phi_2(s) \d^{\a \b}&\equiv \frac 2{\sqrt 3}\bra{K^\a K^\b} X \ket{0}\\
  \end{aligned}
  \label{eq:1}
\end{equation}
where $X$ is any of the operators appearing in (\ref{ff_Gamma})-(\ref{ff_theta}). The relative factor $2 / {\sqrt 3}$ is due to the normalization of the isospin zero eigenstates
\be
| \pi \pi \ra = \f{1} {\sqrt{3}} \sum_{a=1}^3 | \pi^a \pi^a \ra, ~~\text{and} ~~ 
| K K \ra = \f{1} {2} \sum_{a=1}^4 | K^\a K^\a \ra.
\ee 
Also we introduce the $T$-matrix via
\begin{equation}
  S_{i j}(s)\equiv\d_{i j} + 2 i T_{i j} (s) \sqrt{\s _{i} (s) \s_{j} (s)} \, \Theta(s-4m_{i}^{2} ) \, \Theta(s-4m_{j}^{2} ),
\label{S-T-rel}
\end{equation}
where $\Theta(s)$ is the Heaviside theta function (representing the opening of the corresponding channel), and the factors $\s_i(s)$
\begin{equation}
\s_{i}(s) \equiv \sqrt{1-\f {4m_{i}^{2}} {s}}
\end{equation}
are responsible for the phase space volume. One can be straightforwardly convinced that the unitarity of the $S$-matrix (which is represented schematically in Fig.~\ref{fig:2chSMatrix}) translates into the following constraint on $T$
\begin{equation}
  \Im T _{k i} (s) = \sum_{j=1}^2T ^{*} _{i j} (s) T_{k j} (s) \s _{j} (s) \, \Theta(s-4m_j^2), ~~ 4 m_{\pi}^{2}<s<
  \Lambda_\text{dat}^2.
  \label{unitarity-Im-S}
\end{equation}
In a complete analogy the following relation between the imaginary part of the form factors~\eqref{eq:1} and the scattering matrix (Fig.~\ref{fig:2chFormFactors}) can be obtained:
\begin{equation}
  \Im \phi _{i} (s) = \sum_{j=1}^2T ^{*} _{i j} (s) \phi_{j} (s) \s _{j} (s) \, \Theta(s-4m_{j}^2 ),~~ 4 m_{\pi}^{2}<s<
  \Lambda_\text{dat}^2,
  \label{unitarity-Im}
\end{equation}
which we rewrite in an equivalent form
\begin{equation}
\phi_{i}(s) = \sum_{j=1}^2 G _{i j} (s) \phi^{*}_{j}(s), ~~ 4 m_{\pi}^{2}<s<
  \Lambda_\text{dat}^2
\label{unitarity-Hermitian}
\end{equation}
upon defining
\begin{equation}
G _{i j} \equiv \d_{i j} + 2 i \, T_{i j} (s) \, \s_{j} (s) \, \Theta(s-4m_{j}^{2} )~~\text{for}~~i,j = 1,2.
\label{G_MO_equation}
\end{equation}
For energies above $\Lambda_{\text{dat}}$ other channels come into play, correspondingly modifying the constraints (in particular extending the sum in (\ref{unitarity-Im-S})-(\ref{unitarity-Hermitian}) to $j>2$).

\begin{figure}[h]
  \centering %
  \includegraphics[height=5cm]{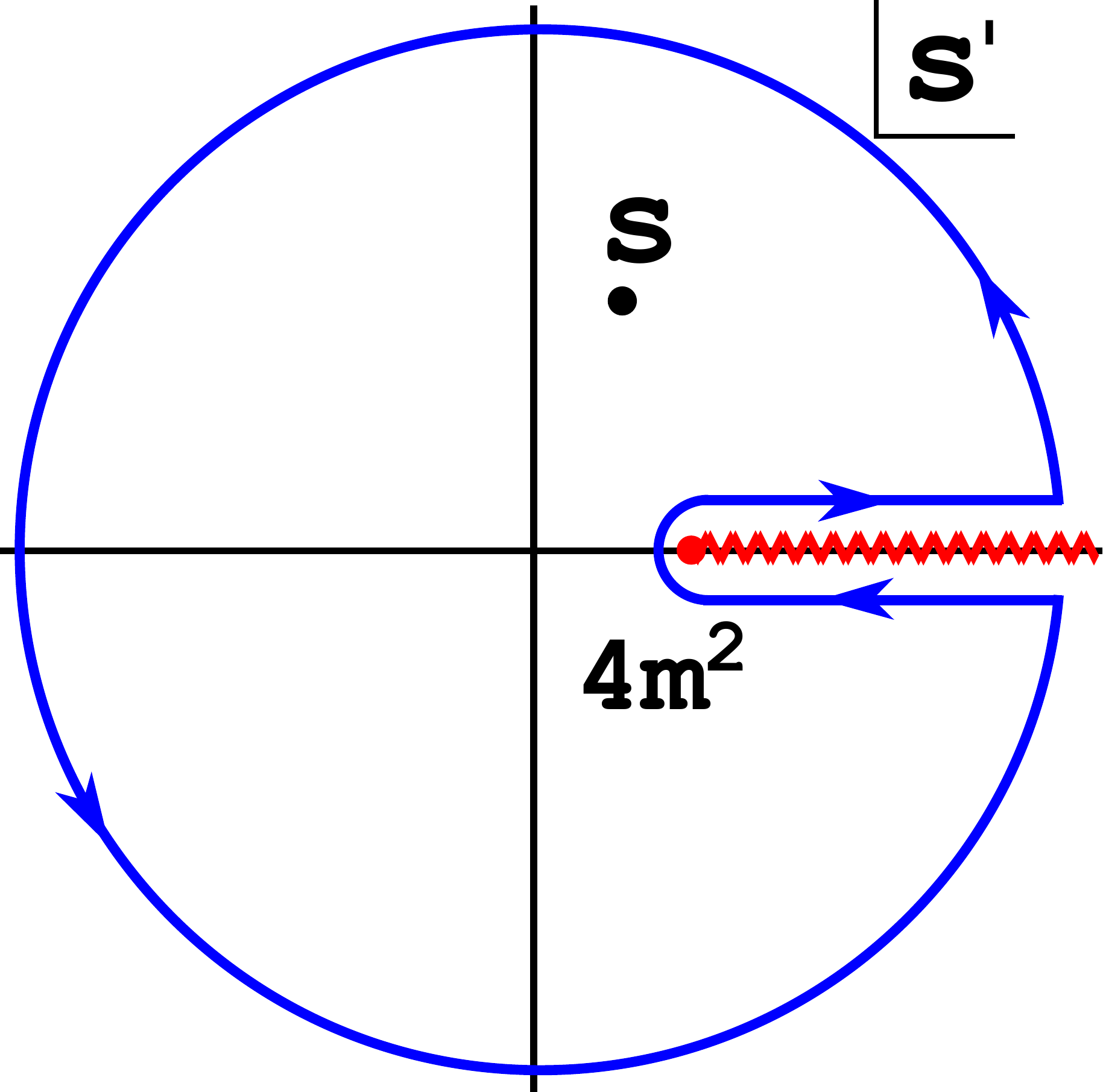}
  \caption[Integration contour]{Integration contour\label{fig:intContour}}
\end{figure}

In deriving the system of equations (\ref{unitarity-Im}) one does not rely on perturbative computations. Therefore, the analogue of the system with all channels taken into account together with analyticity encapsulate all the necessary non-perturbative information about the form factors. In the next Section we will discuss several cases when the solution of (\ref{unitarity-Im}) can be found.

\section{Review of different methods and their results}
\label{sec:review-methods}

The system (\ref{unitarity-Im}) is similar (though not exactly the same) to a more general Hilbert problem of finding a holomorphic vector function $\phi_i(s)$ of finite degree at infinity with a specific  boundary condition (discontinuity) on the cut(s). It is proven (see~\cite{Omnes:1958hv} and the book \cite{Muskhelishvili_book})\footnote{For a more recent presentation see~\cite{Moussallam:1999aq,Yndurain:2003vk,Ananthanarayan:2004xy,Yndurain:2005cm,Oller:2007xd}.} that the number of linearly independent canonical (having no zeros at finite points) solutions, denoted by $\Omega_i^{(1)}(s)$, $\Omega_i^{(1)}(s)$, $\dots$, coincides with the number of channels. The general solution is the represented as a linear combination of the canonical ones
\be
\phi(s) = P_1(s) \Omega^{(1)} + P_2(s) \Omega^{(2)}+ \dots,
\ee
with $P_i(s)$ being polynomials. Moreover, the degree at infinity of the function
\be
D(s) = \det_{ij} \Omega_i^{(j)}(s)
\label{eq:detOmega}
\ee
is completely fixed by the asymptotic behavior of the corresponding $S$-matrix (see a comment after the Eq. (\ref{eq:OmegaAsymp})). However, only in a limited number of cases (specific form of discontinuity, or in our case of the $S$-matrix) the solution can be found. Below we describe several such cases, first showing how the solution can be derived in general and then using the real data for each of them.

\begin{figure}[!t]
\bc
\includegraphics[width=5cm]{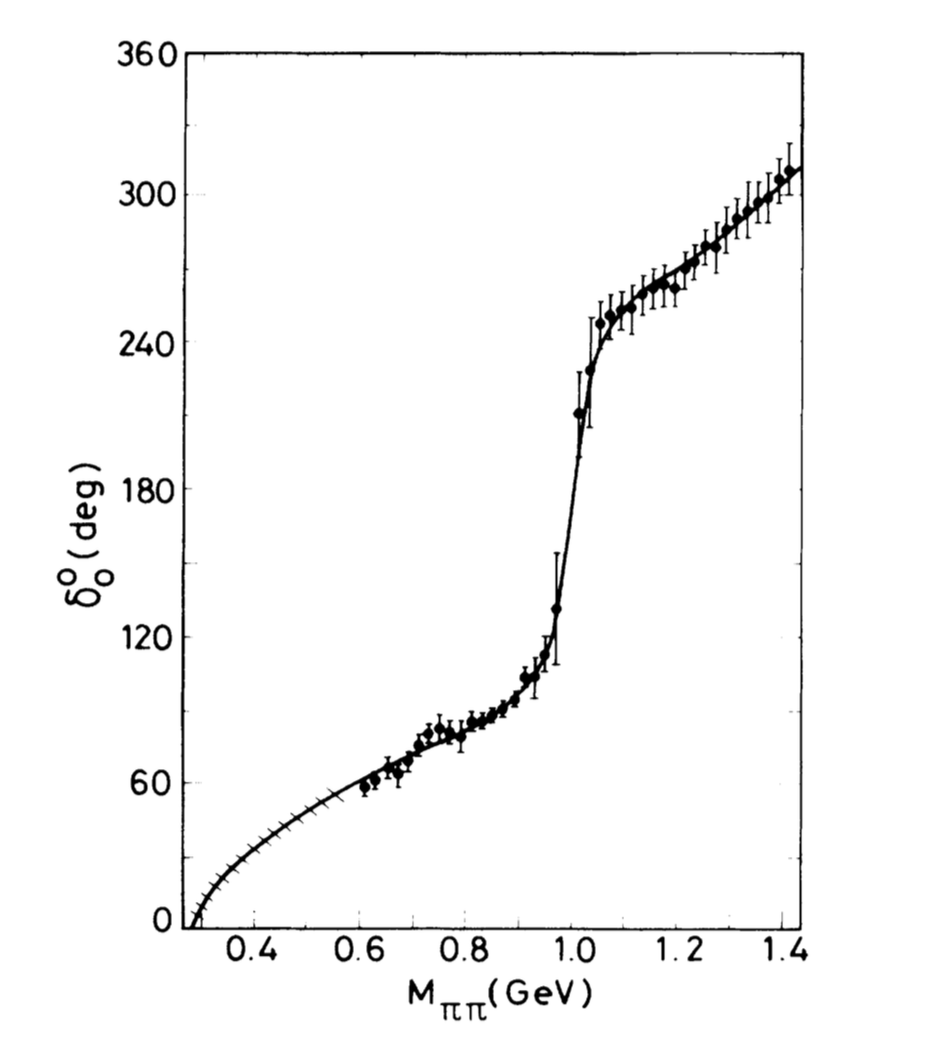}
\caption{\label{pipi_scattering_phase} Pion-pion scattering phase from~\cite{Au:1986vs}.}
\ec
\end{figure}

\subsection{One-channel solution}
\label{sec:1ch}

The first example when the solution can be found explicitly is a hypothetical case of only one channel. As was discussed above, the $S$-matrix in this case is completely specified by just one phase and the system of equations~\eqref{unitarity-Im} reduces to only one equation
\begin{equation}
  \label{eq:8}
  \phi(s) = e^{2 i \delta(s)} \phi^*(s), ~~ s \in \mathbb R.
\end{equation}
According to the rules of analytic continuation the relation above should be satisfied in the whole complex plane. Assuming that the function 
$\phi(s)$ has zeros at $s_{1},s_{2},\dots,s_{N}$, it is easy to show that the function
\begin{equation}
\Omega(s) = \f {\phi(s)} {\prod _{i=1}^{N} (s-s_{i})},
\label{eq:canZero}
\end{equation}
has obviously no zeros and satisfies the same equation (\ref{eq:8}).
As a result
\begin{equation}
  \Im \log \Omega(s) = \d(s),  ~~ s \in \mathbb R,
  \label{dispersion-imaginary-log}
\end{equation}
and the normalized $\Omega(0)=1$ canonical solution, which in this case is called the the Omn\`es factor, can be easily reconstructed by integrating $\log \Omega(s)$ with one subtraction along the contour in Fig.~\ref{fig:intContour}
\begin{equation}
\f{\log \Omega (s)}{s} = \f {1} {\pi} \int _{4m^2}^{\infty} \f{d s'}{s'} \, \f{ \d (s')} {s' - s - i\eps},
\end{equation}
or equivalently
\begin{equation}
\Omega (s) = \exp \l [ \f {s} {\pi} \int _{4m^2}^{\infty} \f{d s'}{s'} \, \f{\d(s')} {s' - s - i\eps} \r].
\label{Omnes-solution-1ch}
\end{equation}
It follows from (\ref{eq:canZero}) that the most general solution (which can have zeros) is given by the product of the Omn\`es factor and a polynomial $P_\phi(s)$ with real coefficients, whose zeros are fixed by that of $\phi (s)$
\begin{equation}
\phi(s) =  P _\phi (s)  \Omega (s).
\label{ff-solution-1ch}
\end{equation}

It is straightforward to show that the asymptotic behavior of the Omn\`es factor is fixed by the scattering phase, appearing in (\ref{Omnes-solution-1ch}), at infinity. Namely,
\be
\Omega(s) \underset{s\to \infty} {\to} s^{-{\d(\infty)} / {\pi}}.
\label{eq:OmegaAsymp}
\ee
It follows from (\ref{S-T-rel}) and (\ref{G_MO_equation}) that in general the determinant (\ref{eq:detOmega}) satisfies the one-channel equation (\ref{eq:8}) with the phase given by $\arg \l ( \det_{ij} S_{ij} \r ) /2 $. Therefore, its asymptotic behavior (see a comment in the beginning of this section) is fixed to be
\be
D(s) \underset{s\to \infty} {\to} s^{-\arg \det_{ij} S_{ij} (\infty) / 2\pi}.
\label{eq:detAsymp}
\ee
The degree at infinity of the determinant is given by the sum of degrees of all canonical solutions. It is obvious for a set of independent one-channel equations, corresponding to the $S$-matrix without mixing between the channels. The determinant in this case is simply a product of Omn\`es factors.

\begin{figure}[h]
 \bc
  \subfloat[\label{fig:piPhase1chCut} One-channel approximation with the cutoff at kaon threshold]{\includegraphics[width=0.45\textwidth]{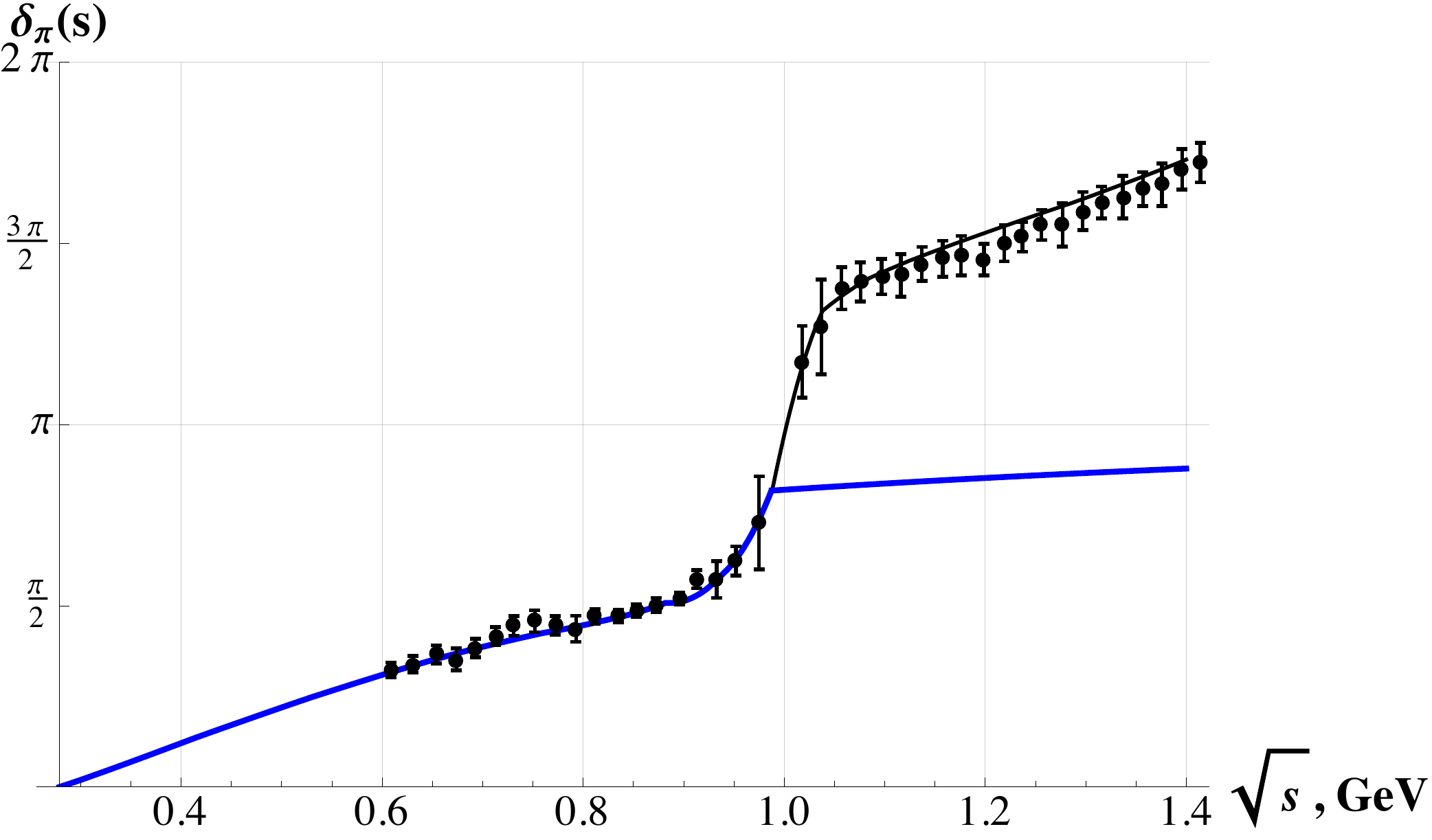}}
  \hfill
 \subfloat[\label{fig:piPhase1chRes} One-channel resonance approximation: $M=0.85$~GeV and $\G = 0.65$~GeV from~\cite{Raby:1988qf}]{\includegraphics[width=0.45\textwidth]{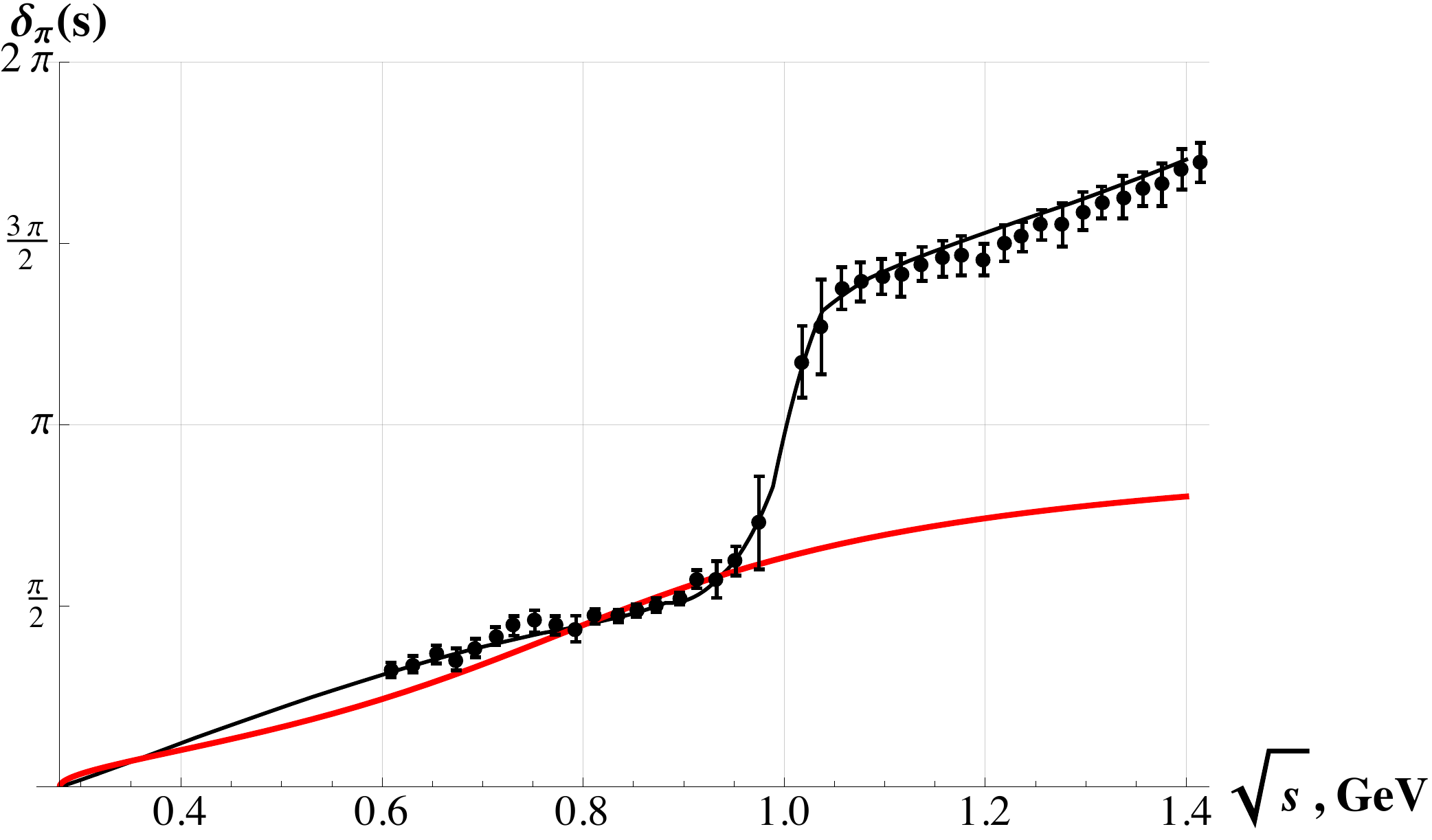}}
  \ec
  \caption{$\pi \pi$ scattering phase shift: black dots correspond to the data from~\cite{Au:1986vs} and black cure is the interpolation. For all approximations the scattering phase is extrapolated to $\pi$ at $s\to \infty$ in a smooth manner.
  \label{fig:piPhase}}
\end{figure}

In order to use the solution (\ref{ff-solution-1ch}) for finding the physical form factors one has to know the scattering phase in (\ref{Omnes-solution-1ch}) and the polynomial $P_\phi(s)$. The region where one-channel approximation could make sense is $s<4m_K^2$. Therefore, the real pion scattering phase (Fig.~\ref{pipi_scattering_phase}) can be used at most up to the kaon threshold. Having no a priory favorable way to extrapolate the phase beyond this point, it is reasonable to extend it asymptotically to the nearest integral value in units of $\pi$ (see Fig.\ref{fig:piPhase1chCut}). The resulting Omn\`es factors $\Omega_{\text{1ch}}(s)$ for different cutoffs are depicted in Fig.~\ref{fig:1chCut1chMod}.

\begin{figure}[h]
\bc
\subfloat[\label{fig:thetaRe1ch} Real part.]{\includegraphics[width=0.45\textwidth]{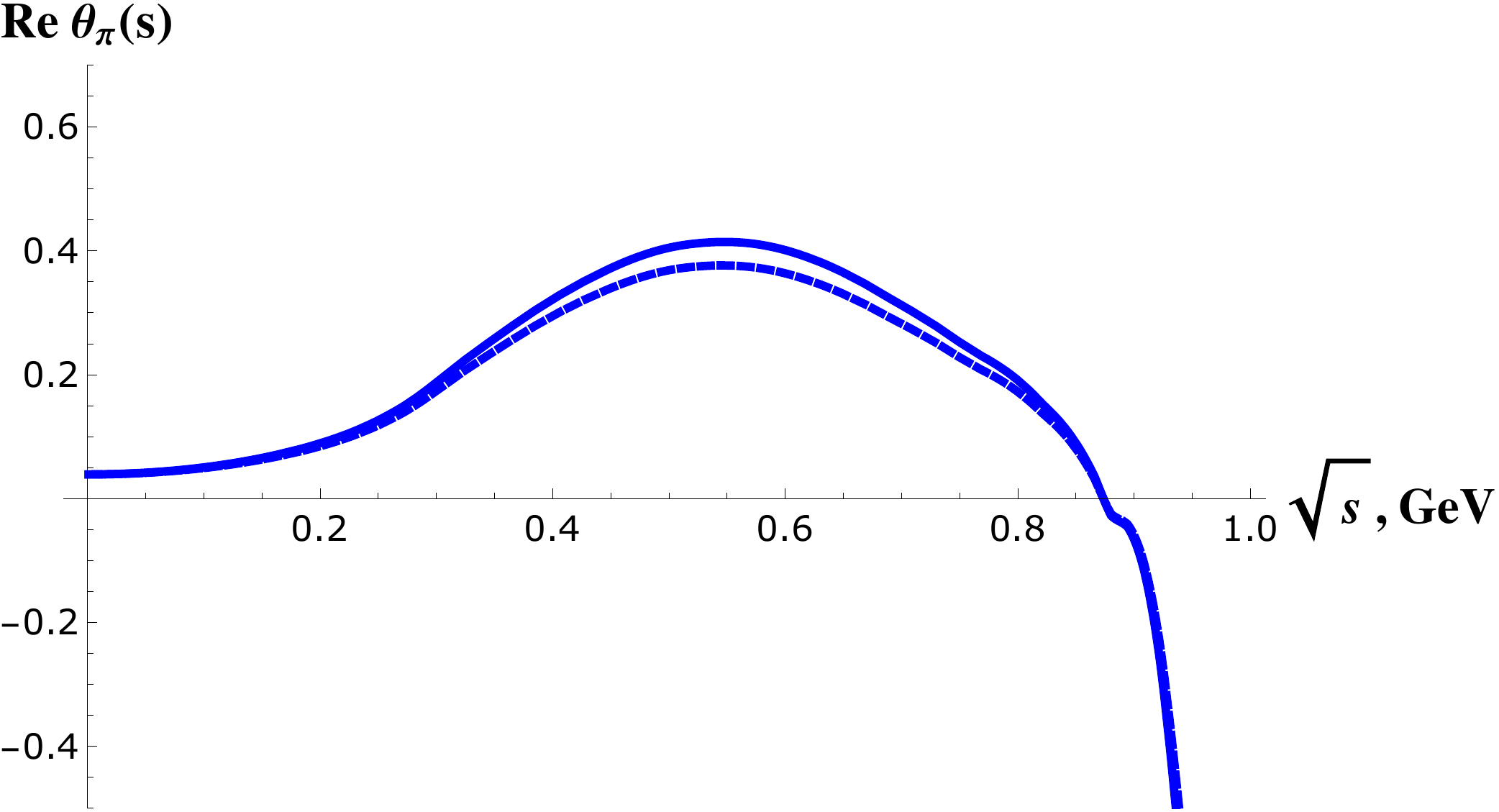}}
\hfill
\subfloat[\label{fig:thetaIm1ch} Imaginary part.]{\includegraphics[width=0.45\textwidth]{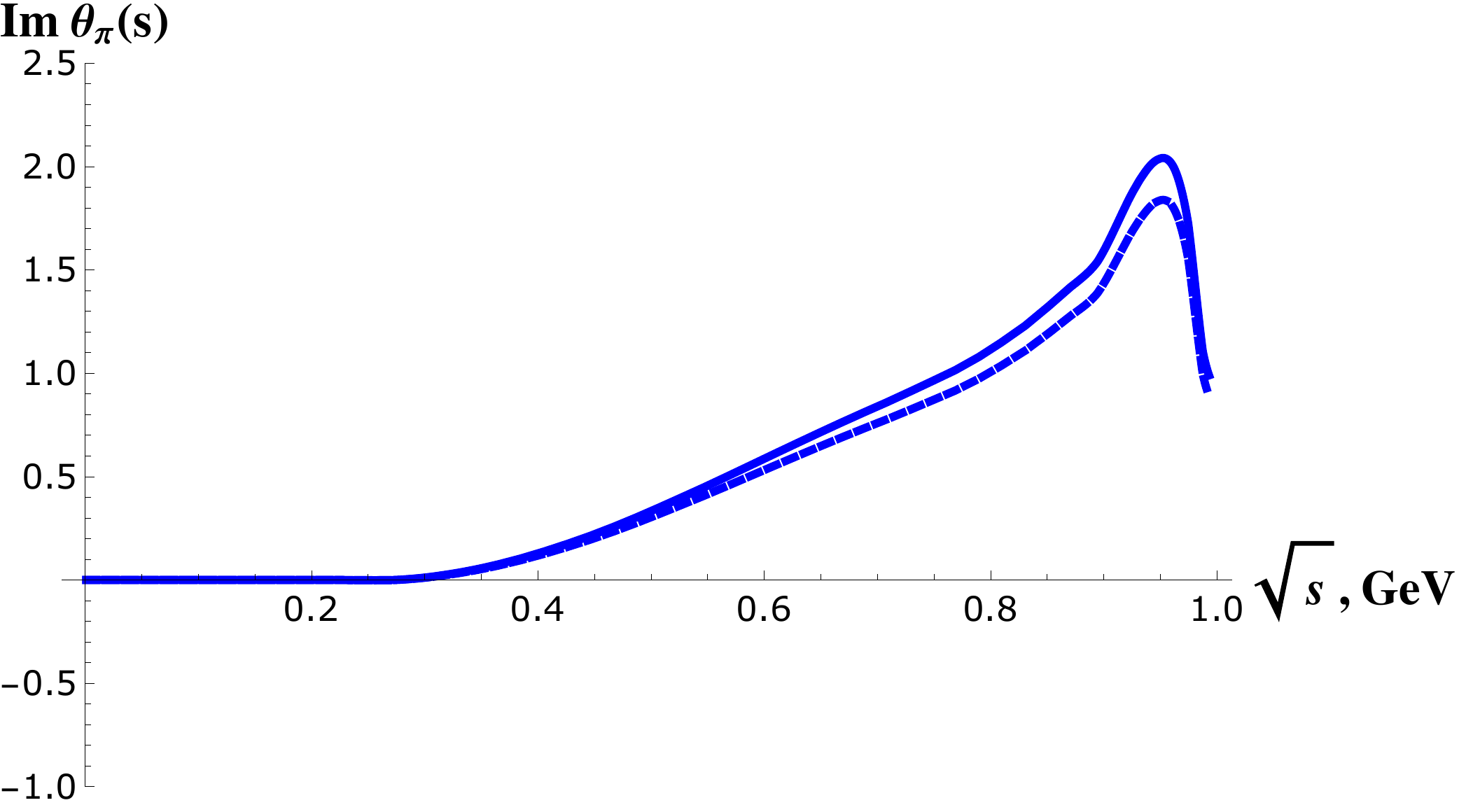}}
\ec
\caption{\label{fig:theta1ch} One-channel approximation result for the form factor $\t_\pi(s)$: solid and dashed lines correspond to boundary conditions (\ref{eq:theta_zero}) and (\ref{eq:theta_prime}) respectively.}
\end{figure}

As for the polynomial $P_\phi(s)$, although it cannot be fixed by the described procedure, there is additional information coming from lattice an ChPT computations, namely, the quadratic scalar radius (\ref{eq:qsrLattice}) and form factors behavior at small energy (\ref{low-energy-theta-ff}), (\ref{low-energy-ff}). The former allows to fix the monomial $as+m_\pi^2$ for $\G_\pi(s)$, by choosing the coefficient $a$ so that the derivative $\G_\pi'(0)$ is consistent with (\ref{eq:qsrLattice}). That results in $a$ being very small, therefore we neglect this term and the we get
\be
\Gamma_\pi^{\text{1ch}}(s) = m_\pi^2 \Omega_{\text{1ch}}(s).
\label{eq:1chG}
\ee
The quadratic scalar radius corresponding to this is $\la r^{2} \ra^\text{1ch}_{S,\pi}=0.62\text{ fm}^{2}.$ There is an ambiguity in implementing a constraint on $\t_\pi(s)$. Indeed, from the expression (\ref{low-energy-theta-ff}) it is clear that the following two conditions are possible
\begin{subequations}
\be
\t_\pi(-2m^2_\pi)=0
\label{eq:theta_zero}
\ee
and
\be
\t_\pi'(0)=1,
\label{eq:theta_prime}
\ee
\end{subequations}
These two differ $m_\pi^2/ \Lambda_\text{QCD}^2$ corrections, and result in the following expressions correspondingly
\be
\theta_\pi^{\text{1ch}} (s) = (s+2m _ \pi^2)\Omega_{\text{1ch}} (s) ~~\text {or} ~~ \theta_\pi^{\text{1ch}} (s) = (a_\text{1ch}s+2m _ \pi^2)\Omega_{\text{1ch}} (s),
\label{eq:1chT}
\ee
with $a_\text{1ch} = 0.9$. In Figure~\ref{fig:theta1ch} we plot both expressions (\ref{eq:1chT}) as functions of $\sqrt{s}$: clearly the curves are very close to each other. 

Lastly, the form factor $\D_\pi(s)$ is given by
\be
\D_\pi^{\text{1ch}} (s) = d_F s \, \Omega_{\text{1ch}} (s),
\label{eq:1chD}
\ee
with $d_F$ a constant (for a way to estimate its value see below). It is clear from (\ref{eq:OmegaAsymp}) that only one of the form factors above has the proper asymptotic $\G_{\text{1ch}}(s)\underset{s\to \infty}{\to}s^{-1}$. This fact testifies that one-channel approximation does not properly describe the underlying dynamics and cannot be complete.

\subsection{Two channels: numerical analysis}
\label{sec:dgl-method}

\begin{figure}[h]
\bc
\subfloat[\label{KK_scattering_phase}Kaon-kaon scattering phase from~\cite{Cohen:1980cq}.]{\includegraphics[width=0.5\textwidth]{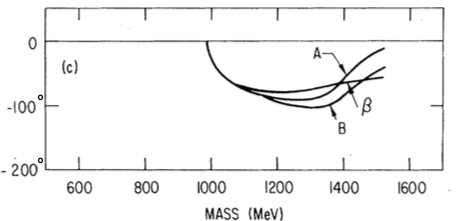}}
\subfloat[\label{elasticity} Elasticity parameter from~\cite{Hyams:1973zf}.]{\raisebox{0.4cm}{\includegraphics[width=0.5\textwidth]{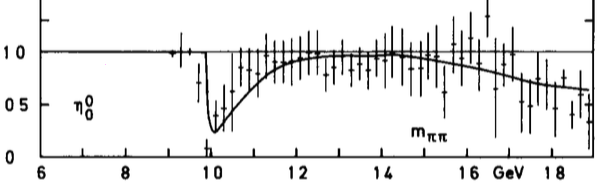}}}
\ec
\caption{Two channel $S$-matrix parameters. \label{fig:phaseElasticity}}
\end{figure}

As was already mentioned an explicit analytic solution to the system (\ref{unitarity-Im}) is generally not available. Therefore, one resorts to numerics. For the two channels the $S$-matrix is a $2\times 2$ unitary matrix
\begin{equation}
S= 
\left ( 
  \begin{array}{ccc}
    \eta e ^{2i \d _{\pi}} & i \sqrt{1-\eta ^{2}} e^{i(\d_{\pi}+\d_{K})}\\
    i \sqrt{1-\eta ^{2}} e^{i(\d_{\pi}+\d_{K})} & \eta e ^{2i \d _{K}}
  \end{array}
\right),
\label{S-matr-T}
\end{equation}
where $\d_{\pi,K}$ are the scattering phases of $\pi\pi\to \pi\pi$ and $\bar K K \to \bar K K$ and $\eta$ is the elasticity parameter, characterizing  the mixing between the two channels. All these parameters are extracted from experimental data (see e.g.~\cite{Hyams:1973zf,Au:1986vs} and Fig.~\ref{fig:phaseElasticity}).

It is precisely the mixing due to the elasticity parameter that precludes one from finding a solution analytically. Indeed, were it not for $\eta(s)\neq 1$, the system of equations (\ref{unitarity-Im}) would factorize into two independent one-channel equations (\ref{eq:8}) that could be solved explicitly. However, one can still derive the analytic form of the determinant (\ref{eq:detOmega}) even for a non-trivial elasticity parameter. Indeed, assuming asymptotic values for the scattering phases $\d_\pi(\infty) =2\pi$, $\d_K(\infty) =0$ it follows immediately from (\ref{eq:detAsymp}) that the determinant of the canonical solutions decays at infinity as $s^{-2}$.

In~\cite{Donoghue:1990xh} 
the following iterative procedure for solving the system was suggested (further developed in~\cite{Moussallam:1999aq,Ananthanarayan:2004xy}). In the zeroth approximation the functions $\phi^{(0)}_1(s)$, $\phi^{(0)}_2(s)$ are initialized by constants: $\phi_1^{(0)}(s) = 1$ and $\phi_2^{(0)} (s)= \lambda \in \mathbb{R}$. The real and imaginary parts of $\phi_i(s)$ at the step $n+1$ are computed via
\begin{equation}
  \Im \phi ^{(n+1)}_i (s)= \Re\l [ \sum_{j}T ^{*} _{i j} (s) \phi ^{(n)}_{j} (s) \s _{j} (s) \, \Theta(s-4m_{j}^2 ) \r],~~s\in \mathbb R,
  \label{eq:10}
\end{equation}
and
\begin{equation}
  \label{eq:9}
\Re \phi ^{(n+1)}_i (s) = \f {1} {\pi} \dashint _{4m_\pi^2}^{\infty} d s' \f{\Im \phi_i ^{(n)}(s')} {s'-s},
\end{equation}
Since the two numerically obtained solutions tend to zero for $s\to \pm \infty$ it is concluded that these are the two canonical solutions
$\Omega_\text{2ch}^{(1)}(s)$ and $\Omega_\text{2ch}^{(2)}(s)$, behaving as $s^{-1}$ at infinity, in agreement with $s^{-2}$ behavior of the determinant discussed above.

Then, the form factors (\ref{ff_Gamma})-(\ref{ff_theta}) are obtained by considering proper linear combinations of
the numerically obtained solutions. With normalized canonical solutions
\be
\Omega_i^{(j)}(0) = \d_i^j ,
\label{eq:2ch_canonical_norm}
\ee
where for brevity we have omitted the subscript (2ch) and the lower index specifies the component of the corresponding solution,
one obtains in general the form factors
\begin{subequations}
\begin{align}
\label{2ch_ff_Gamma}
\G_\pi^\text{2ch}(s) & = m_\pi^2 \Omega_1^{(1)} + \f {2}{\sqrt{3}} \G_K (0) \Omega_1^{(2)} \\
\label{2ch_ff_Delta}
\D^\text{2ch}_\pi(s) &= \f {2}{\sqrt{3}} \D_K (0) \Omega_1^{(2)} , \\
\label{2ch_ff_theta}
\t^\text{2ch} _\pi(s) &= (2m_\pi^2+ p s ) \Omega_1^{(1)} + \f {2}{\sqrt{3}} \l ( \t_K(0) + q s \r ) \Omega_1^{(2)}, 
\end{align}
\label{2ch_ff}
\end{subequations}
with coefficient $p$ and $q$ are related to slopes of $\t'_{\pi, K}(0)$. In~\cite{Donoghue:1990xh} the slope of pion form factor is taken to be $\t'_{\pi}(0)=1$ and other unknown parameters in (\ref{2ch_ff}) are obtained using $SU(3)\times SU(3)$ ChPT, which is not extremely reliable. In particular the following values were chosen
\be
\G_K(0) = \f {m_\pi^2} {2}, ~~ \D_K(0) = m_K^2 - \f {m_\pi^2} {2}, ~~ \t_K(0) = 2m_K^2, ~~ \t'_{K}(0)=0.9-1.1.
\label{eq:ff_K_bc}
\ee
As we can see the polynomial needed to fix the low energy behavior of the form factor (\ref{2ch_ff_theta}) has degree one, so this form factor does not have the proper asymptotic at infinity which is again the signal of incompleteness of the approach (see the discussion in~\cite{Donoghue:1990xh}).

We would like to bring to readers' attention that while the numerical procedure produces two decaying at infinity solutions, it has not been proven that so obtained solutions are the canonical ones. There may not even exist canonical solutions all decaying at infinity, it is only the determinant (morally speaking the product) that has a specific asymptotic behavior. It is obvious if the limit $\eta(s)=1$ is considered. In that case the two canonical solutions, according to asymptotic of pion and kaon scattering phases, behave as $s^{-2}$ and $s^{0}$ correspondingly. By construction the method gives access to form factors only for $s\in \mathbb R$, hence it is not obvious how to check that no additional singularities (and zeros) have been generated and the two solutions are analytic in the whole complex plane with the cut.

\subsection{Resonance approximation \label{sec:resonance}}

A class of $T$-matrices for which explicit solutions can be constructed is the following. Taking into account the equation (\ref{S-T-rel}) we observe immediately that
\be
\phi_i (s)= \sum_j c_j(s) T _{j i} (s),
\label{eq:5}
\ee
with $c_k(s)$ being real functions on $s\in \mathbb R$, formally solves the system (\ref{unitarity-Im-S}). Generally, $T$-matrix has a left hand cut\footnote{The right hand cut $s\geq 4m^2$ corresponds to $u\leq 0$ for $t=0$ (s-channel), which due to the crossing implies the existence of the left-hand cut $s \leq 0$.} and the expression (\ref{eq:5}) cannot represent any of the form factors. However, if there exists an analytic function $T(s)$ with only the right-hand cut and approximating properly the scattering data (phase shifts and elasticities), then the formula (\ref{eq:5}) provides us with a proper solution to (\ref{unitarity-Im-S}).

\begin{figure}[t]
\bc
\subfloat[\label{fig:2chResPiPhase} $\pi\pi$ scattering phase]
{\includegraphics[width=0.3\textwidth]{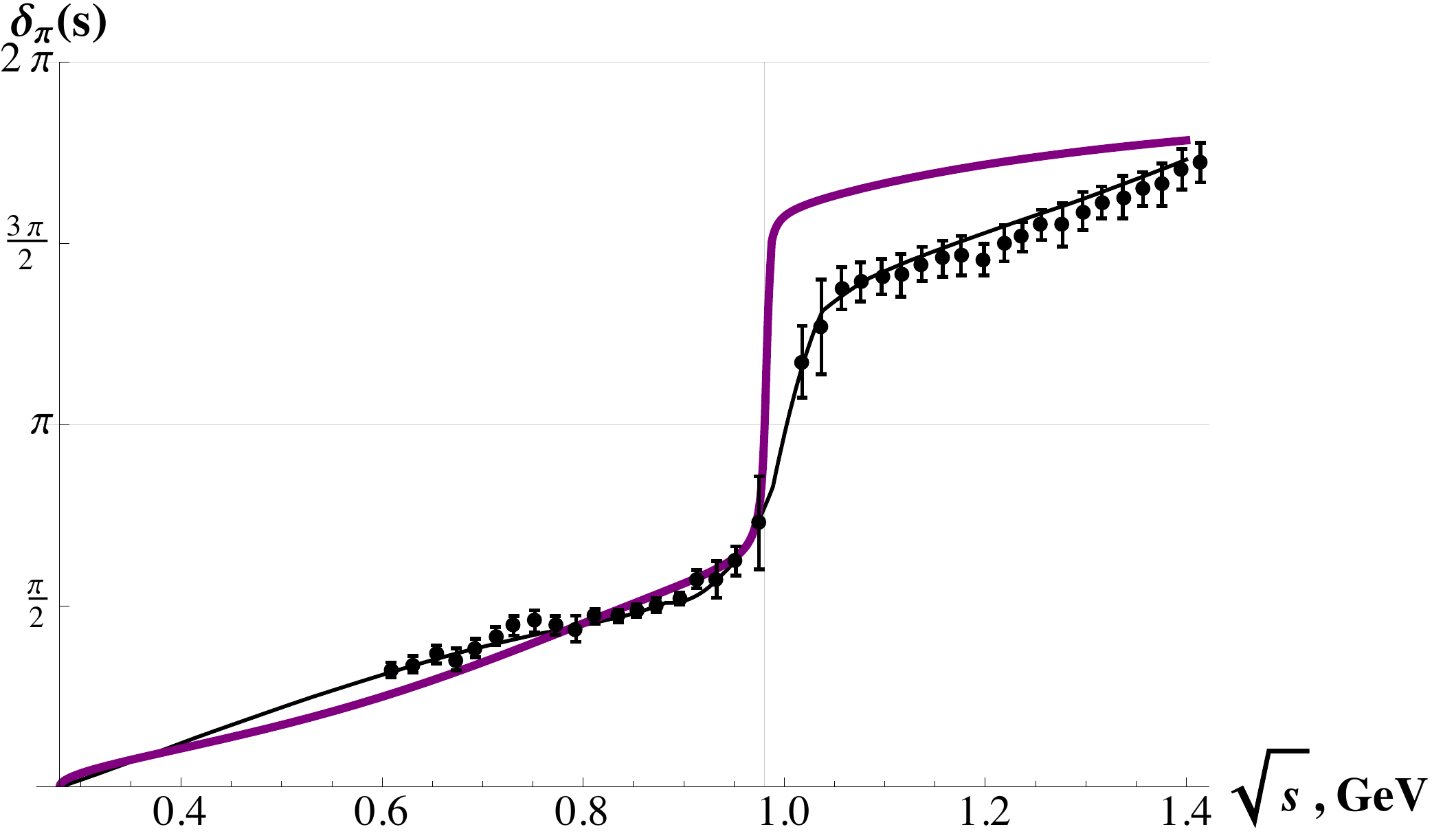}}
\subfloat[\label{fig:2chResKPhase} $\bar K K$ scattering phase]
{\includegraphics[width=0.3\textwidth]{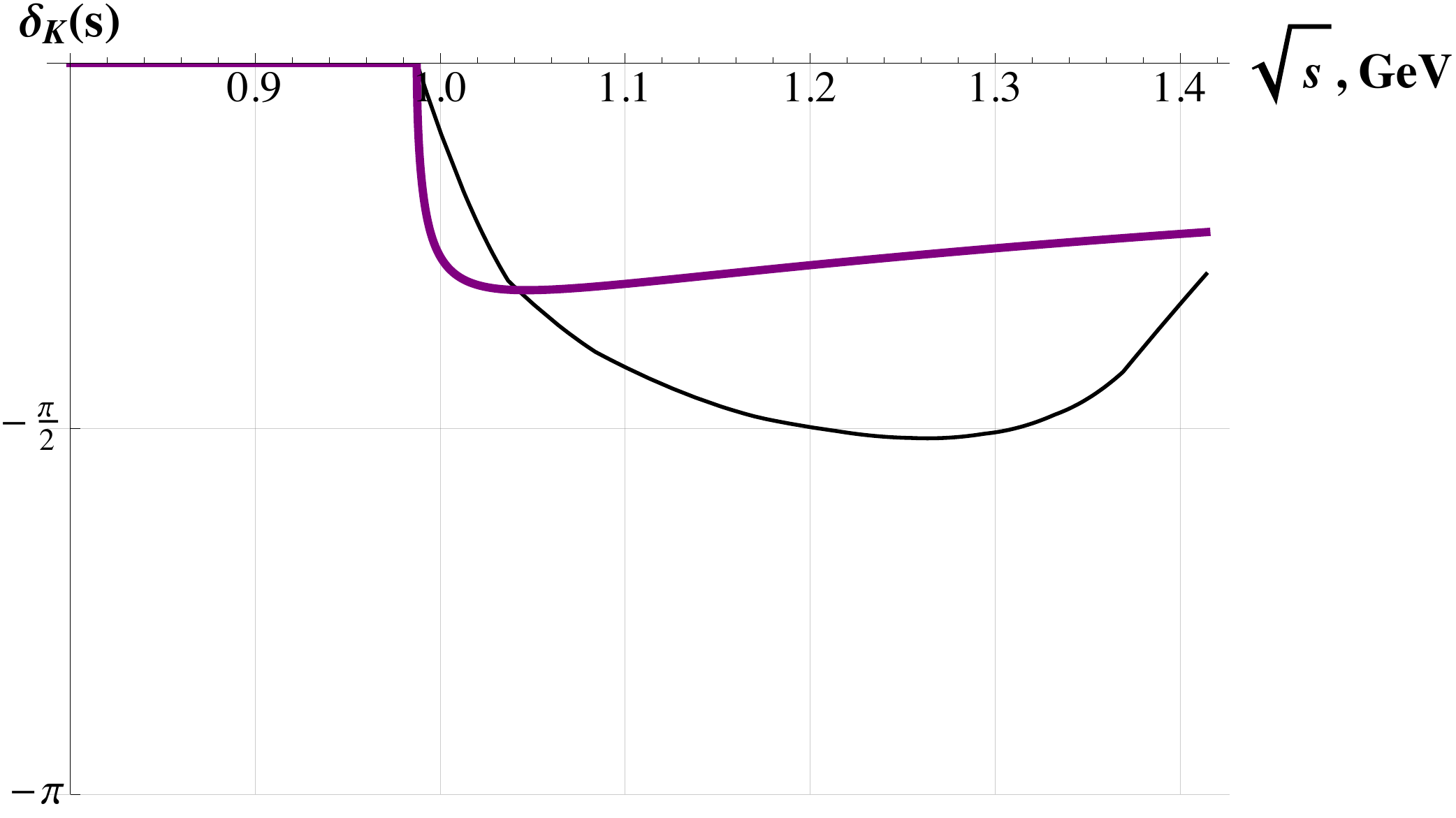}}
\subfloat[\label{fig:2chResElasticity} Elasticity]
{\includegraphics[width=0.3\textwidth]{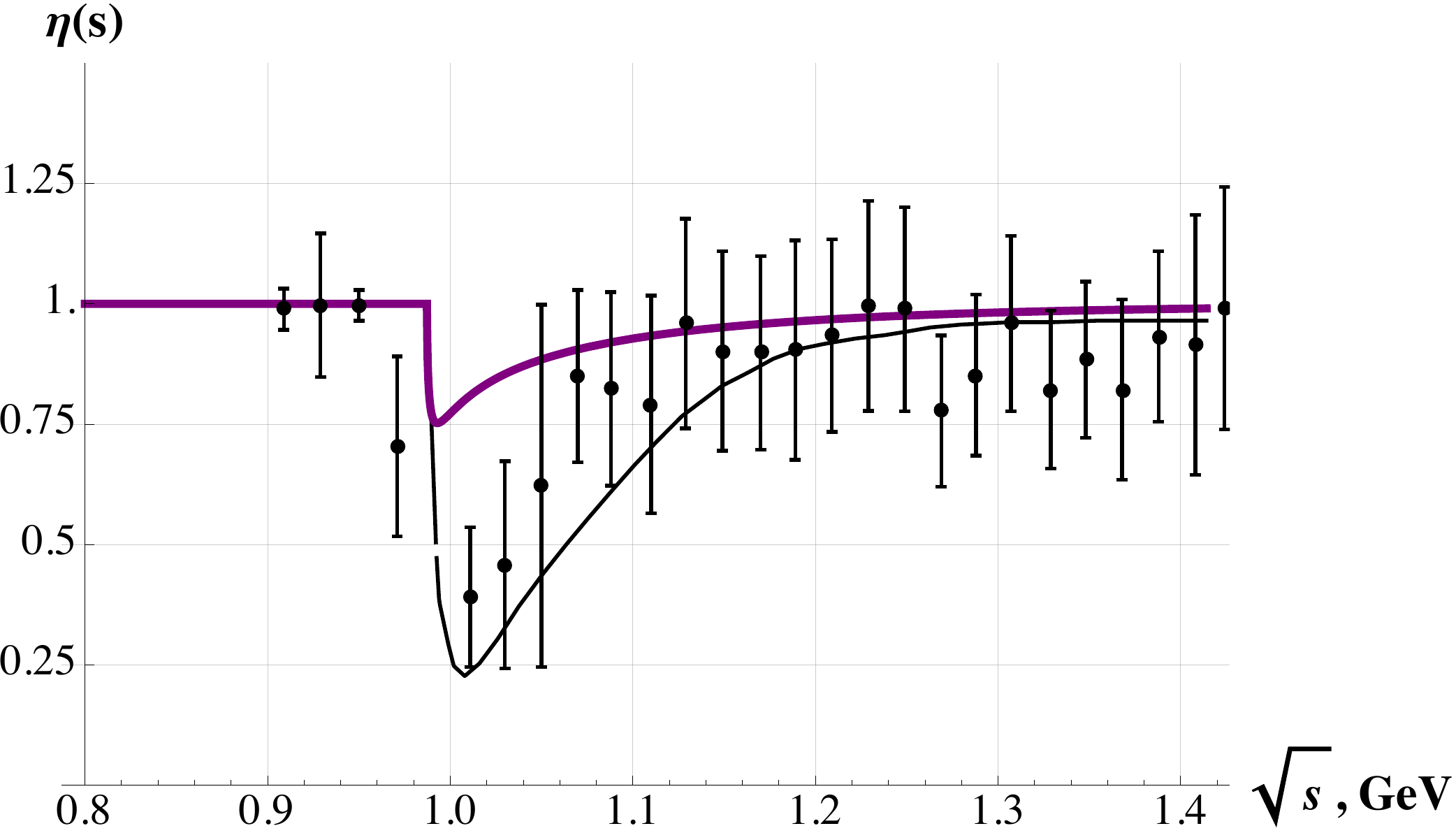}}
\ec
\caption{\label{fig:2chRes} Two-channel resonance approximation for the $S$-matrix with the following values for the parameters $M_1=0.87$~GeV, $\G_1=0.7$~GeV, $M_2=0.92$~GeV, $\G_2=1$~GeV and $\lambda=-0.2$ from~\cite{Truong:1989my} }
\end{figure}

The function considered in~\cite{Raby:1988qf} and~\cite{Truong:1989my} has the resonance form, e.g. there are poles. However, these are, as usual, restricted to the second (unphysical) sheet. The function is built in the following way. Assuming that $T$ is symmetric one can show using (\ref{unitarity-Im-S}) that in matrix notations the following relation is satisfied
\be
\mathrm {Im} \, \bm T ^{-1} = - \bm \s, ~~ s \in \mathbb R
\ee
with $\bm \s = \mathrm {diag} \l ( \s_1(s), \s_2(s) \dots \r )$. It follows from the above equation that the $T$-matrix can be formally written as
\be
\bm T = (\bm A - i \bm \s)^{-1},  ~~ s \in \mathbb R,
\ee
with $\bm A$ an arbitrary real-valued (on $\mathbb R$) matrix.

Both in~\cite{Raby:1988qf} and~\cite{Truong:1989my} the simplest choice was made
\be
A_{i j} (s) = \f{P_{ij}^{(1)}(s)} {\sqrt{s}},
\label{eq:resonanceChoice}
\ee
with $P_{ij}^{(1)}(s)$ being first order polynomials. The functions $c_k(s)$ were also chosen to be linear, with coefficients fixed by the low energy behavior of form factors (\ref{low-energy-theta-ff}) and (\ref{low-energy-ff}).

In particular for the one-channel $T$-matrix resonance representation
\begin{equation}
  \label{eq:6}
  T^{-1} (s) = \f {M^2 - s + \Gamma \sqrt{4m_\pi^2-s}}{\sqrt{s} \Gamma},
\end{equation}
the normalized $\Omega_{\text{res1}}(0)=1$ solution was found in~\cite{Raby:1988qf}\footnote{There is a factor $2$ difference in the notation for the width between our notation and the one used in the paper cited.} 
\be
\Omega _ {\text{res1}} (s) = \f {M^2+2m_\pi \G} {M^2 - s + \Gamma \sqrt{4m_\pi^2-s}},
\label{eq:OmegaRes}
\ee
with the following parameters for mass and width of the resonance $M=0.8$~GeV and $\G = 0.65$~GeV. Note that with these parameters it is straightforward to compute the quadratic scalar radius of the pion, using that
\be
\G_\pi^\text{res1}(s) = m_\pi^2 \Omega_\pi^\text{res1}(s),
\ee
which results in $\la r^{2} \ra_{S,\pi} = 0.56$~fm$^2$. Meanwhile to reproduce the value (\ref{eq:qsrLattice}) one takes $M=0.85$~GeV and $\G = 0.8$~GeV. The other two form factors are taken to be
\be
\D^\text{res1}_\pi(s) = 0, ~~ \t^\text{res1}_\pi(s) = (s+2m_\pi^2)\Omega^\text{res1}_\pi(s).
\ee

For two channel resonance approximation the $T$-matrix was chosen to have the form~\cite{Truong:1989my}
\begin{equation}
  \label{eq:7}
  \bm T^{-1} = \left(
    \begin{array}{ccc}
      \dst \f {M_1^2 - s + \Gamma_1 \sqrt{4m_\pi^2-s}}{\sqrt{s} \Gamma_1}& - \lambda \sqrt{s} \\
      - \lambda \sqrt{s} & \dst \f {M_2^2 - s + \Gamma_2 \sqrt{4m_K^2-s}}{\sqrt{s} \Gamma_2}
 \end{array}
  \right),
\end{equation}
with constants  $M_1=0.87$~GeV, $\G_1=0.7$~GeV, $M_2=0.92$~GeV, $\G_2=1$~GeV now describing positions and widths of two resonances and $\lambda=-0.2$ being responsible for the coupling between the two channels\footnote{Our notations are slightly different from the ones used in the paper cited.}. Knowing the form of the $T$-matrix, one builds canonical solutions normalized as in (\ref{eq:2ch_canonical_norm}), and then the form factors according to (\ref{2ch_ff})\footnote{The authors of~\cite{Truong:1989my} follow a different procedure. They use ChPT result directly for the amplitude (\ref{LO-apm}) to impose the corresponding boundary conditions. This is effectively leads to $\D_\pi(s)=0$.}.

\begin{figure}[h]
\bc
\subfloat[\label{fig:Gamma1chRes2chRes} Form factor $\G_\pi(s)$. One-channel resonance approximation (red), two-channel resonance approximation with $\G_K(0)=0$ and $\G_K(0)=2m^2_\pi$ (purple solid and dashed respectively).]{\includegraphics[width=0.45\textwidth]{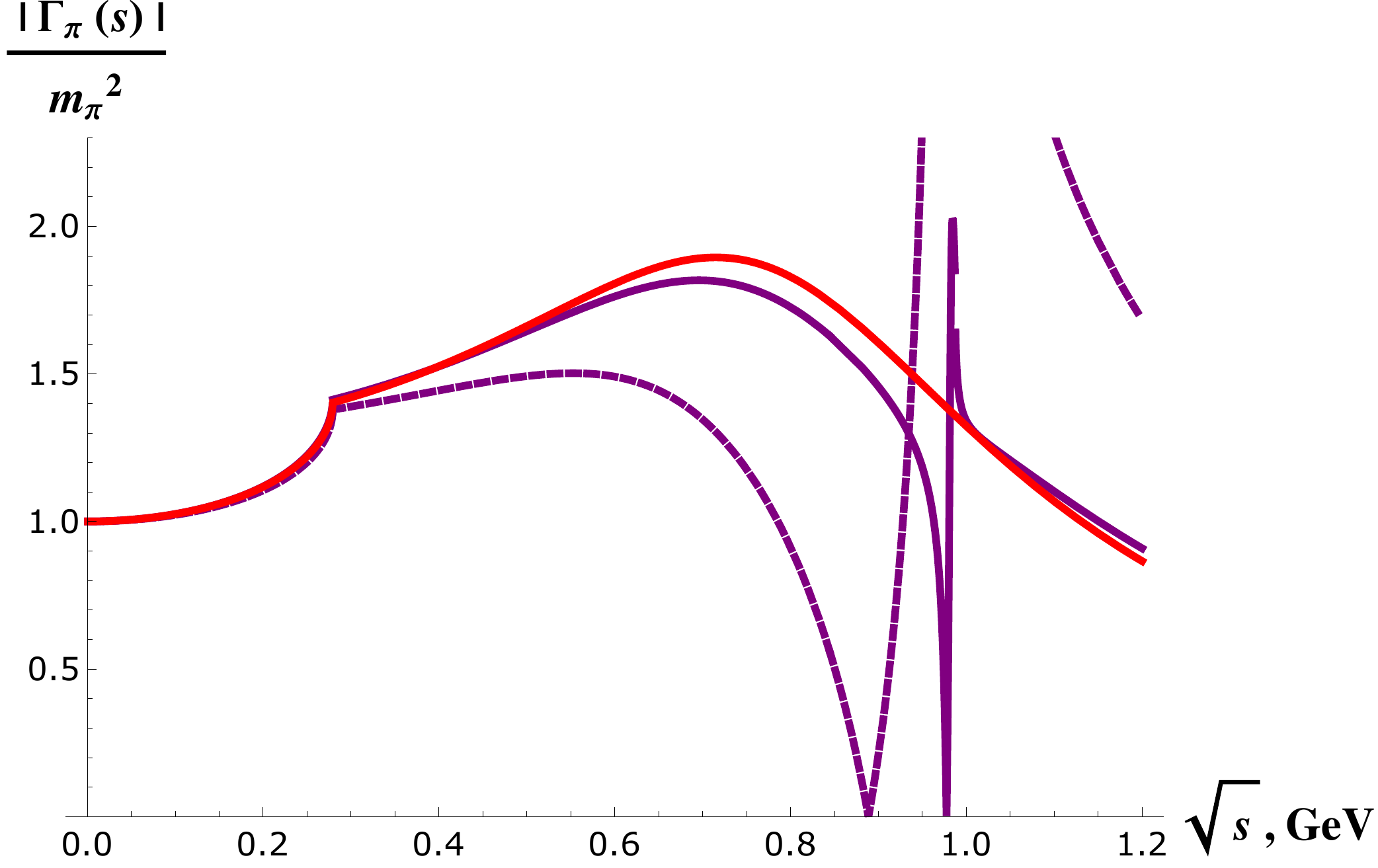}}
\hfill
\subfloat[\label{fig:Delta2chRes} Form factor $\D_\pi(s)$. Two-channel resonance approximation with $\D_K(0)$ differing from the one chosen in (\ref{eq:ff_K_bc}) by $+30\%$ (dashed) and $-30\%$ (solid).]{\includegraphics[width=0.45\textwidth]{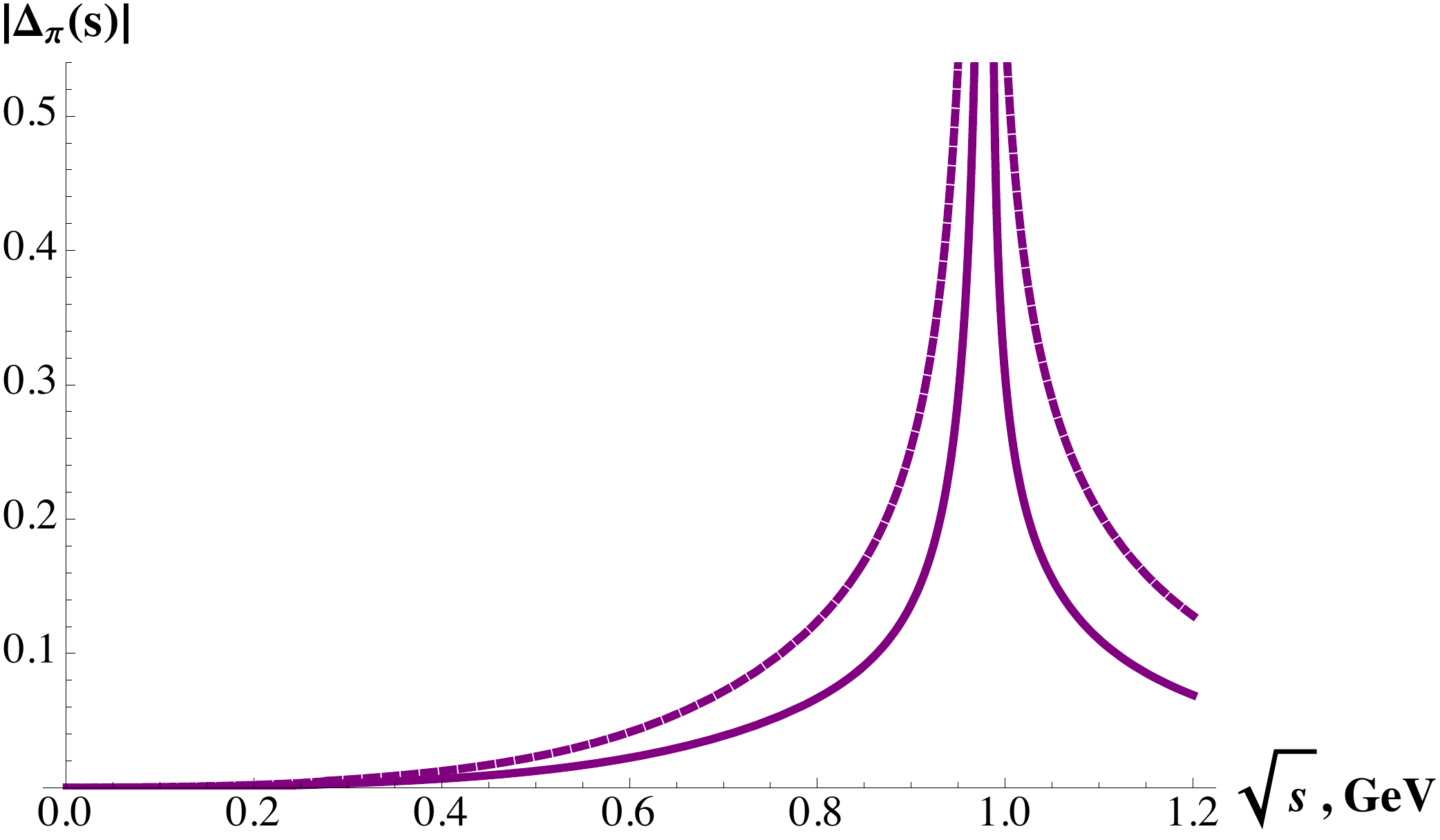}}
\ec
\caption{\label{fig:2chRes_analysis} Form factors obtained with two-channel (\ref{2ch_ff}) resonance approximations with different parameters.}
\end{figure}

Clearly, the form (\ref{eq:resonanceChoice}) of the matrix $\bm A$ is not unique and the corresponding solution~\eqref{eq:5} 
can be used to approximate real form factors works inasmuch as the ansatz~\eqref{eq:6} and \eqref{eq:7} fits the experimental data well. It is evident from Fig.~\ref{fig:piPhase1chRes} that the one-channel resonance approximation does not account for data points above kaon threshold. For the two-channel resonance approximation of~\cite{Truong:1989my}, depicted in Fig.~\ref{fig:Delta2chRes}, the situation is only somewhat better. 

Therefore, there is no good reason to expect that the resonance approximation can give quantitatively reliable approximations for the form factors. However, since the scattering data is qualitatively reproduced, it is a good model to test other techniques and assumptions. In particular we would like to test how sensitive the results are with respect to variations of the parameters in (\ref{2ch_ff}). It can be checked that the form factor $\t_\pi(s)$ is almost insensitive to those variations. At the same time as evidenced in Fig.~\ref{fig:2chRes_analysis}, form factors $\G_\pi(s)$ and $\D_\pi(s)$ depend substantially on the values of $\G_K(0)$ and $\D_K(0)$.

\subsection{Modified one-channel approximation}
\label{sec:modified-one-channel}

In this section we present an approximate solution to the Muskhelishvili-Omn\`es problem.
It is based on a specific feature of the data, namely, the fact that the elasticity parameter $\eta (s)$ appearing in the definition of the S-matrix (\ref{S-matr-T}) is consistent with $\eta = 1$ almost everywhere apart from $2m_K \lesssim \sqrt{s}\lesssim \unit[1.1]{GeV}$ ~\cite{Cohen:1980cq,Kaminski:1997rr,Guerrero:1998ei} (see Fig.~\ref{elasticity}).
This in turn implies that the equation~\eqref{eq:8}, i.e. 
\be
\phi_\pi (s) = e^{2 i \delta_ \pi (s)} \phi_\pi^*(s),
\ee
is satisfied almost everywhere except for a narrow window just above the kaon threshold. We therefore take the scalar form factor $\phi_\pi(s)$ in the form (\ref{Omnes-solution-1ch}) with all the experimental data for the pion scattering phase and not only up to kaon threshold as for the one-channel approximation discussed in the Section~\ref{sec:1ch}. Above $\sqrt{s}=\Lambda_{\text{dat}}$ we extrapolate the phase in a smooth manner to $2\pi$, which now is the nearest integral value in units of $\pi$. We call this a modified one-channel approximation. It has been previously discussed in~\cite{Yndurain:2003vk,Ananthanarayan:2004xy,Yndurain:2005cm,Caprini:2005an,Yndurain:2005gb,Oller:2007xd}.

As before we use ChPT and (\ref{eq:qsrLattice}) to fix polynomials defining the form factors. As a result we get
\begin{subequations}
\begin{align}
\label{eq:1chModG}
\Gamma_\pi^{\text{mod}}(s) & = m_\pi^2 \f {s_\G-s} {s_\G} \Omega_{\text{mod}}(s), \\
\label{eq:1chModD}
\D_\pi^{\text{mod}} (s) & = d_F s \, \Omega_{\text{mod}} (s), \\
\label{eq:1chModT}
\theta_\pi^{\text{mod}} (s) & = (s+2m _ \pi^2)\Omega_{\text{mod}} (s),
\end{align}
\label{eq:1chModFF}
\end{subequations}
where $d_F$ is a yet unfixed constant and $\sqrt{s_\G}=1.1$~GeV.

As we discussed in the Section~\ref{sec:1ch} once the canonical solution is known all other solutions are parametrized by a polynomial. Therefore, as a cross check the method of this section should reproduce the result of the numerical procedure from the Section~\ref{sec:dgl-method}, otherwise it would not even qualify as a valid approximation. From the Figs.~\ref{fig:1chCut1chMod}, \ref{fig:thetaFF2} and \ref{fig:DeltaFF} we see that this is precisely what happens. The only subtlety is that in order to account for the mismatch in asymptotic behaviors of 
$\t_\pi^\text{DGL}(s)$ and $\t_\pi^\text{mod}(s)$, it is necessary to multiply the latter by a linear function 
\be
\t_\pi^\text{mod} (s) \to \f{s_\t-s}{s_\t} \, \t_\pi^\text{mod} (s) \approx \t_\pi^\text{DGL}(s),
\label{eq:addZero}
\ee 
with a free parameter $s_\t=1.3$~GeV$^2$.

\section{Results and comparison}
\label{sec:results}

\begin{figure}[t]
\bc
\subfloat[\label{fig:1chRes} One- and two-channel resonance approximation with parameters from~\cite{Raby:1988qf} and~\cite{Truong:1989my} with $\G_K(0)=0$ correspondingly (red, purple).]
{\includegraphics[width=0.45\textwidth]{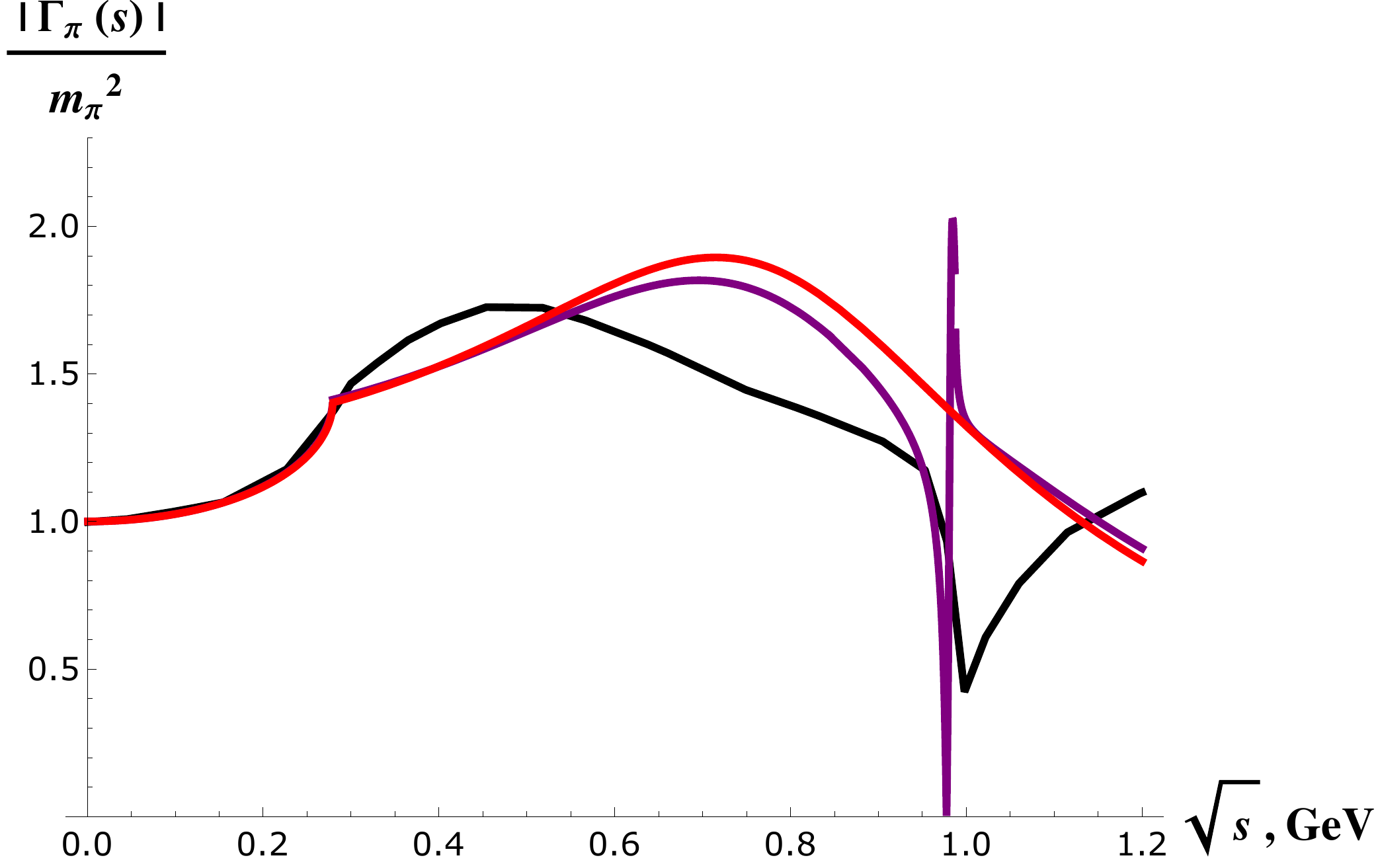}}
\hfill
\subfloat[\label{fig:1chCut1chMod} One- and modified one-channel approximations with $\sqrt{s_\G} \approx 1$~GeV and $\sqrt{s_\G} = 1.1$~GeV (blue, solid and dashed green).]
{\includegraphics[width=0.45\textwidth]{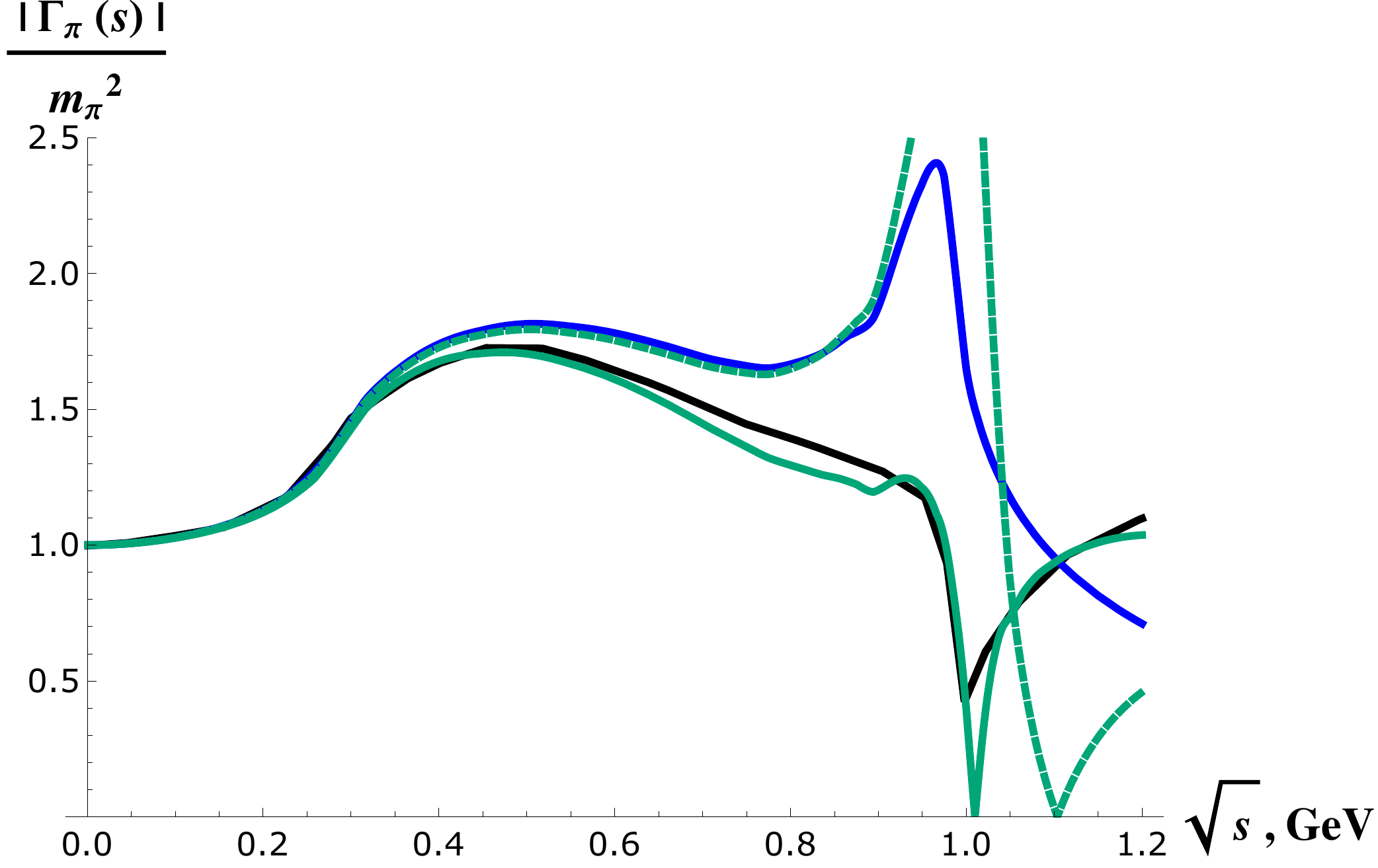}}
\ec
\caption{\label{fig:GammaFF} Form factor $\G_\pi(s)$ in units of $m_\pi^2$
  obtained using different methods. In both plots the black curve is the same,
  taken from~\cite{Ananthanarayan:2004xy}. }
\end{figure}

In this section we compare results for the form factors and corresponding decay rates obtained using different methods. In Fig.~\ref{fig:GammaFF} we plot the absolute value of the form factor $\Gamma_\pi(s)$ in units of $m_\pi^2$. It is evident that all methods are numerically consistent up to $s=4m_\pi^2$. All methods except one-channel approximation exhibit a characteristic dip in the vicinity of $\sqrt{s} = 1$~GeV. It could be argued that all the results are comparable up to $\sqrt{s}\approx 0.6-0.7$~GeV and disagree above. Also it is clear from (\ref{fig:1chCut1chMod}) that even a slight variations of the parameter $s_\G$ of the modified one-channel approximation, causing insignificant changes in the slope at $s=0$, or equivalently in the quadratic scalar radius of the pion\footnote{The parameter $\sqrt{s_\G} = 1.1$~GeV corresponds to the central value of (\ref{eq:qsrLattice}), while $\sqrt{s_\G} = 1$~GeV, effectively reproducing the result of~\cite{Ananthanarayan:2004xy}, leads to $\la r^{2} \ra_{S,\pi}=0.57 \text{ fm}^{2}$, which is at the edge of the error interval.}, lead to the dramatic difference around $\sqrt{s} = 1$~GeV.

\begin{figure}[h]
\bc
\subfloat[\label{fig:thetaReMod}  Real part]{\includegraphics[width=0.45\textwidth]{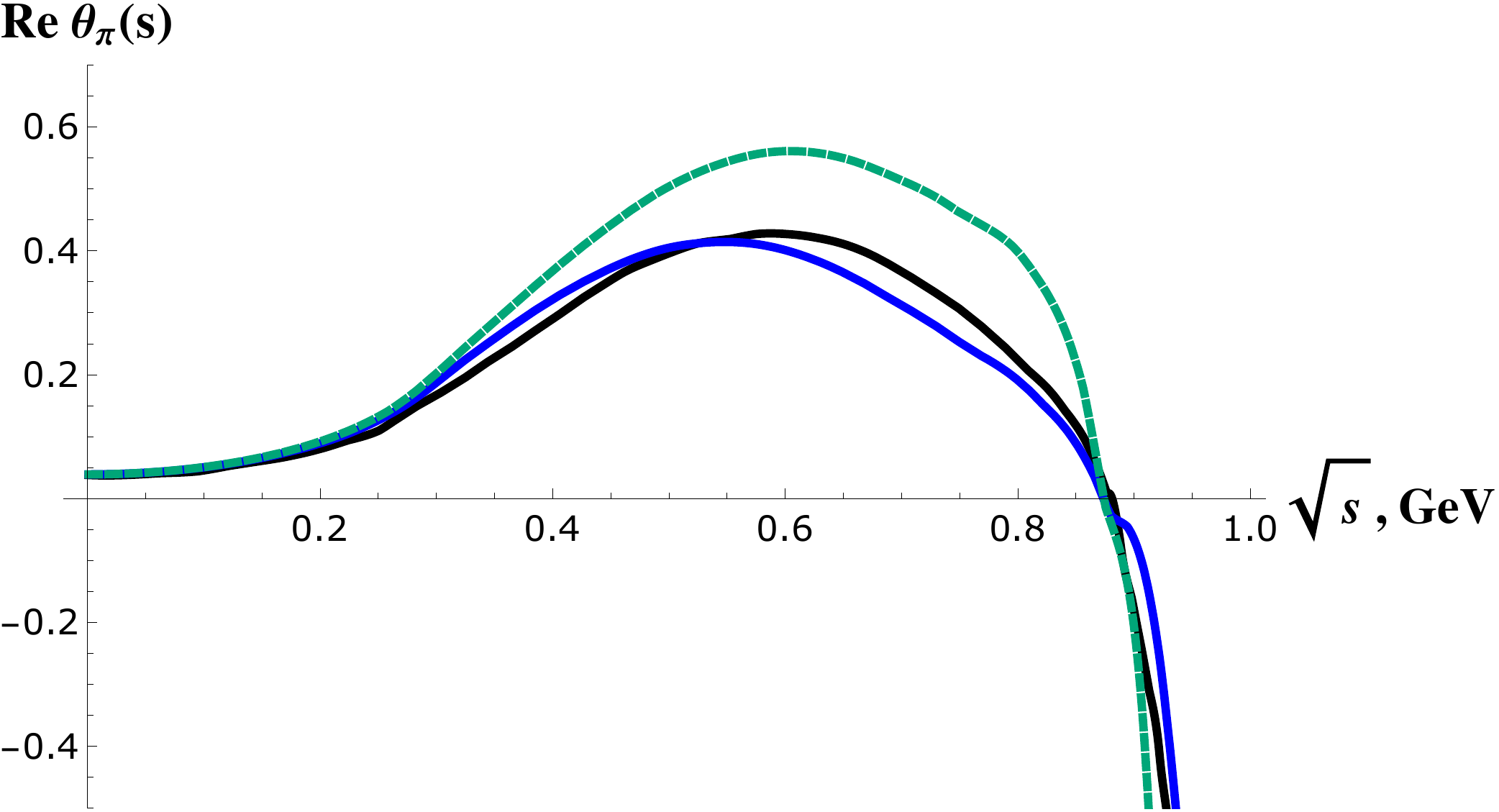}}
\hfill
\subfloat[\label{fig:thetaImMod} Imaginary part.]{\includegraphics[width=0.45\textwidth]{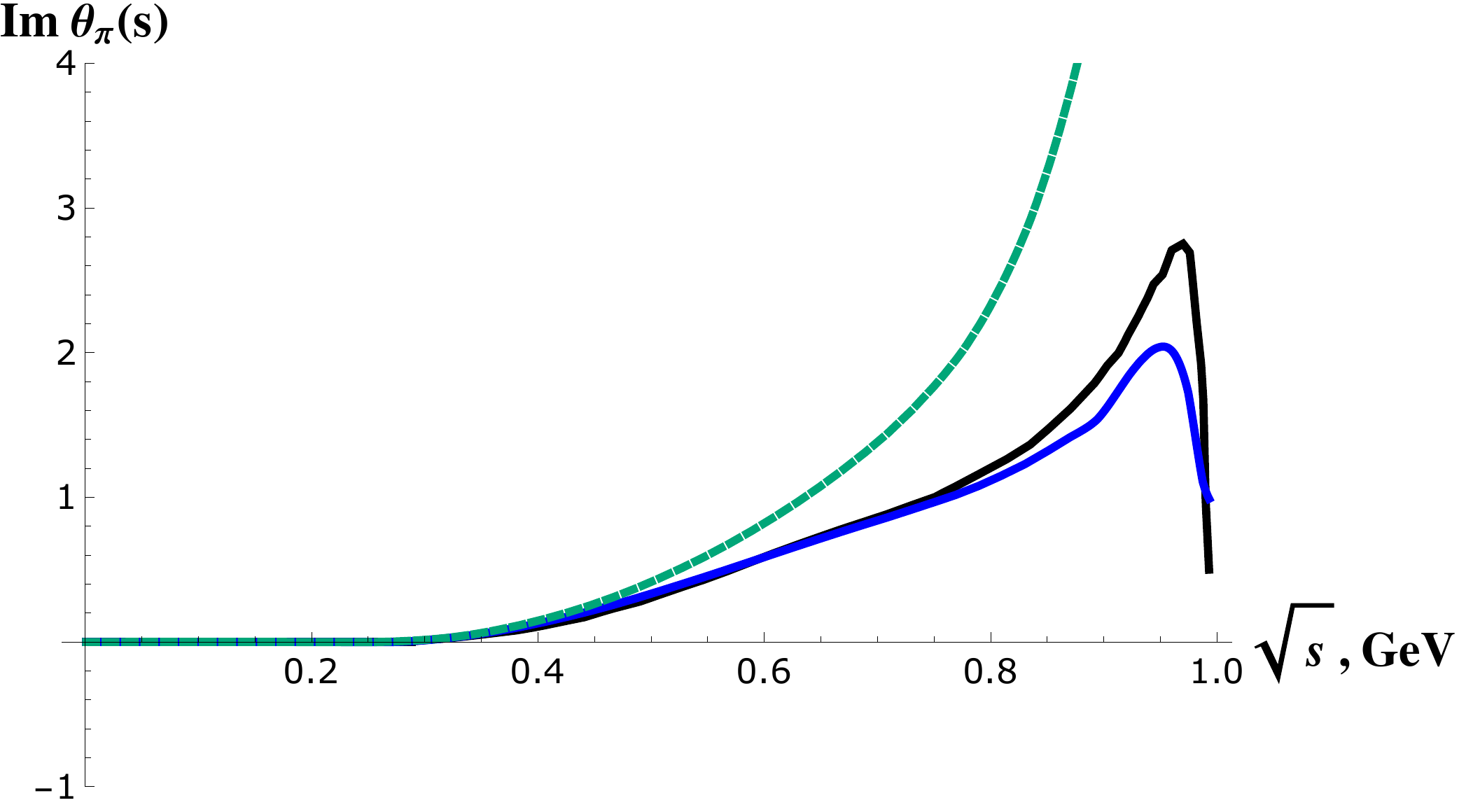}}
\ec
\caption{\label{fig:thetaFF} Real and imaginary parts of the form factor $\t_\pi(s)$ for different methods: one-channel approximation (blue), modified one-channel approximation (green), black curve is taken from~\cite{Donoghue:1990xh}.}
\end{figure}

\begin{figure}[h]
\bc
\subfloat[\label{fig:thetaReMod+0} Real part.]{\includegraphics[width=0.45\textwidth]{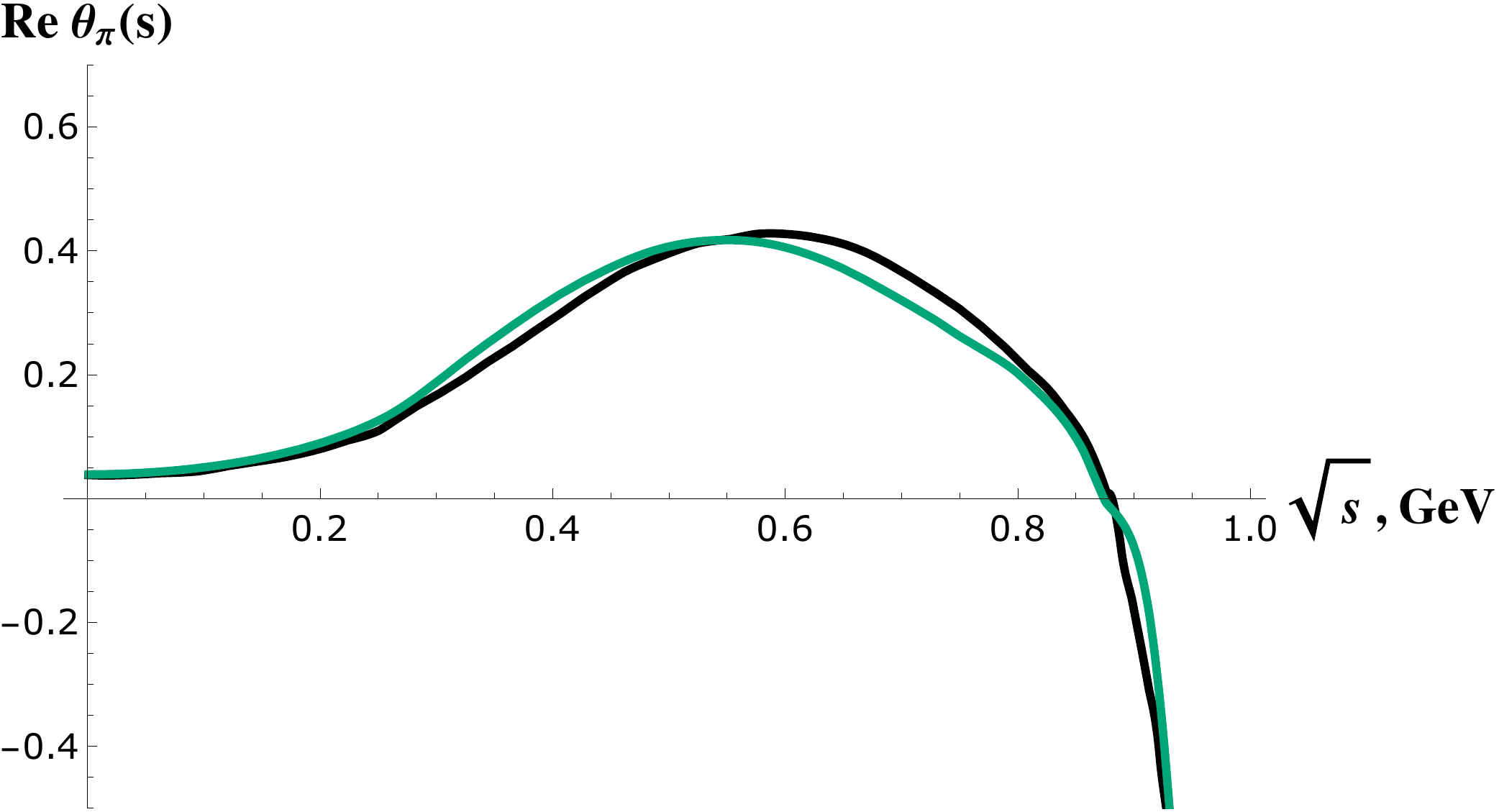}}
\hfill
\subfloat[\label{fig:thetaImMod+0} Imaginary part.]{\includegraphics[width=0.45\textwidth]{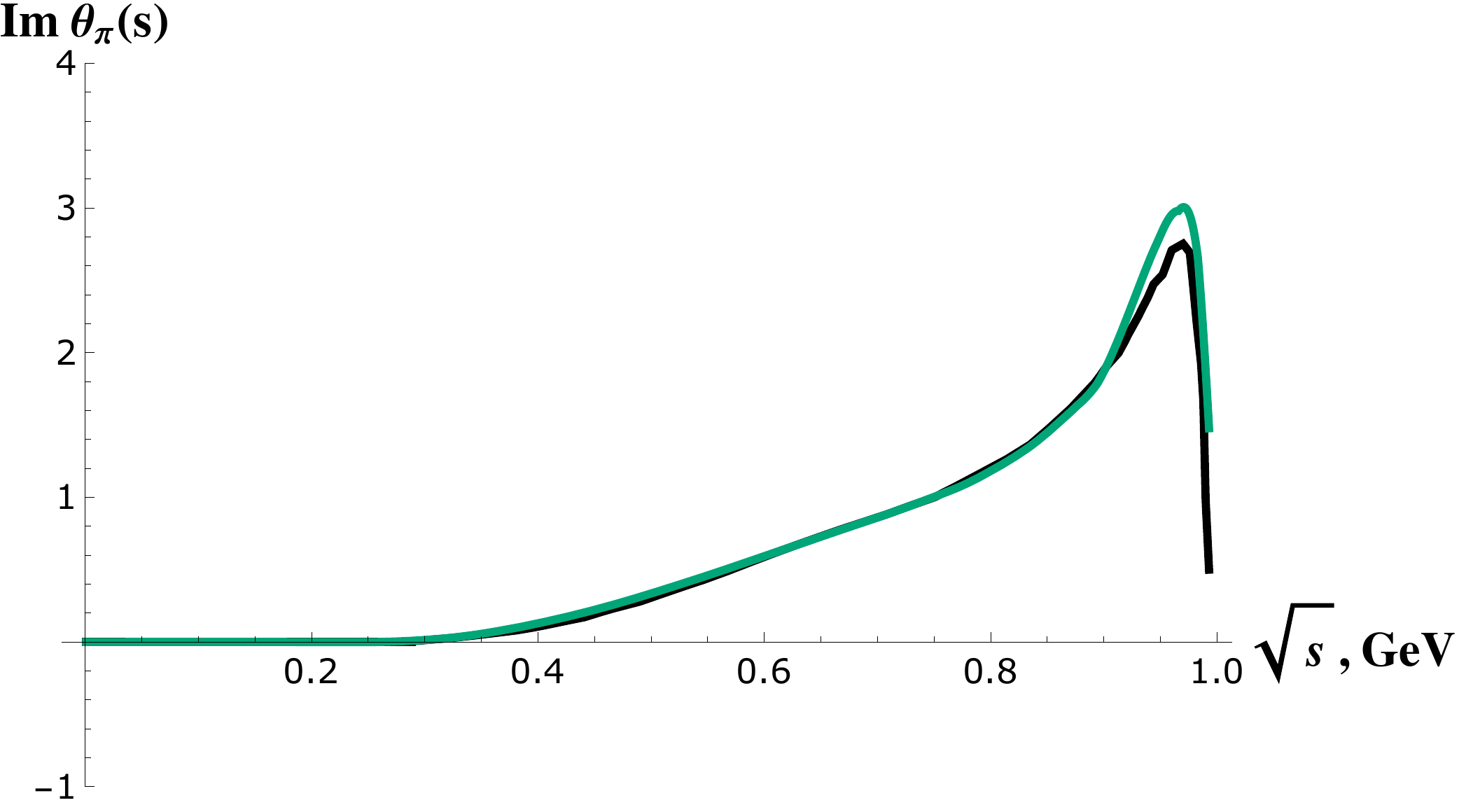}}
\ec
\caption{\label{fig:thetaFF2} Reproducing the result for the form factor $\t_\pi(s)$ with the modified one-channel approximation by using an additional zero (\ref{eq:addZero}) with $s_\t=1.3$~GeV$^2$.}
\end{figure}

In Fig.~\ref{fig:thetaFF} we plot both the real and the imaginary parts of the form factor $\t_\pi(s)$. We see that the one-channel approximation produces the result close to that of the two-channel numerical treatment, while the deviation of the modified one-channel approximation becomes clear rather early. The reason is that the combination that reproduces the result of~\cite{Donoghue:1990xh} is given in (\ref{eq:addZero}) and not the form factor $\t_\pi^\text{mod}(s)$ itself. In Fig.~\ref{fig:thetaFF2} we plot the combination (\ref{eq:addZero}) with $\sqrt{s_\t} = 1.1$~GeV.

\begin{figure}[h]
\bc
\includegraphics[width=9cm]{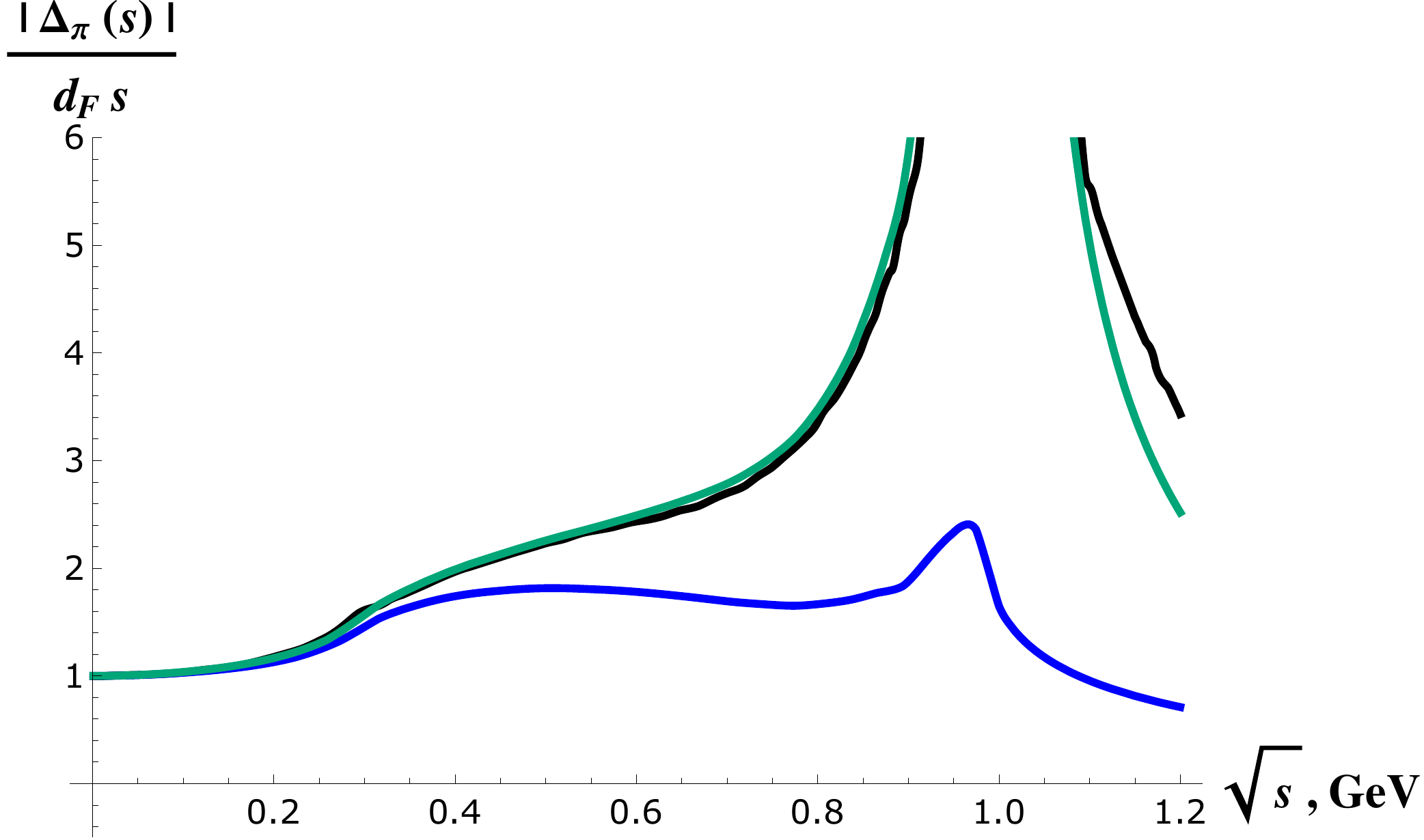}
\ec
\caption{\label{fig:DeltaFF} Absolute value of the form factor $\D_\pi(s)$ in units of $d_F s$: one-channel approximation (blue), the result for the two-channel numerical procedure from~\cite{Ananthanarayan:2004xy} (black), the modified one channel approximation (\ref{eq:1chModD}) (green) coincides with the numerical procedure up to $\sqrt{s}\approx 1$~GeV.
}
\end{figure}

In Fig.~\ref{fig:DeltaFF} we plot the form factor $\D_\pi(s)$. The modified one channel approximation again passes the cross checking for it reproduces the numerical two-channel result of~\cite{Donoghue:1990xh,Ananthanarayan:2004xy}. On the other hand, the one-channel approximation is clearly inadequate when applied to that specific form factor. Contrary to $\G_\pi(s)$ and $\t_\pi(s)$ where the one channel approximation could be argued to -- at least qualitatively -- coincide with the two-channel analysis up to $\sqrt{s}=0.6$~GeV, in the case of 
$\D_\pi(s)$  it fails already at rather small values of the center of mass energy. A possible reason for that can be the presence of a relatively broad $f_0(980)$ resonance very close to kaon threshold, which is not captured at all by the one-channel approximation, for it 
does not take into account the real data above $4m_K^2$. The disproportionate change of the form factor $\D_\pi(s)$ when compared to that of $\Gamma_\pi(s)$ and $\theta_\pi(s)$ indicates an enhanced relative significance of the $s$-quark effects for the process at hand.

\begin{figure}[h]
\bc
\subfloat[\label{fig:1chMod} The one-channel (blue) and the modified one-channel approximation with $s_\t=1.3$~GeV$^2$, $ s_\G = 1.2$~GeV$^2$ (solid green) and $s_\t=1.2$~GeV$^2$, $s_\G=1$~GeV$^2$ (dashed green).]{\includegraphics[width=0.45\textwidth]{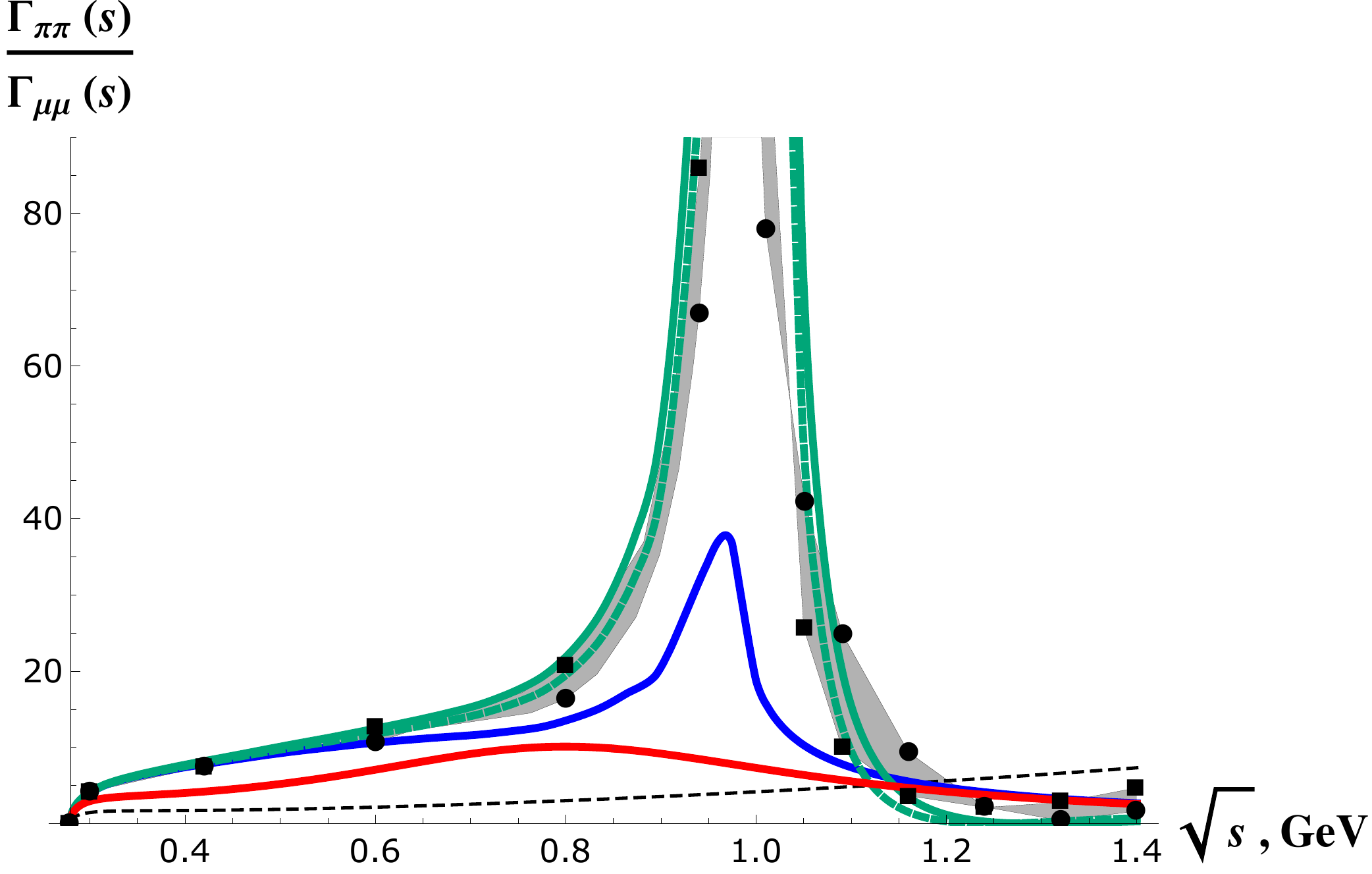}}
\hfill
\subfloat[\label{fig:1chRes2chResDGL} Two-channel resonance approximation with parameters from~\cite{Truong:1989my} (purple).]{\includegraphics[width=0.45\textwidth]{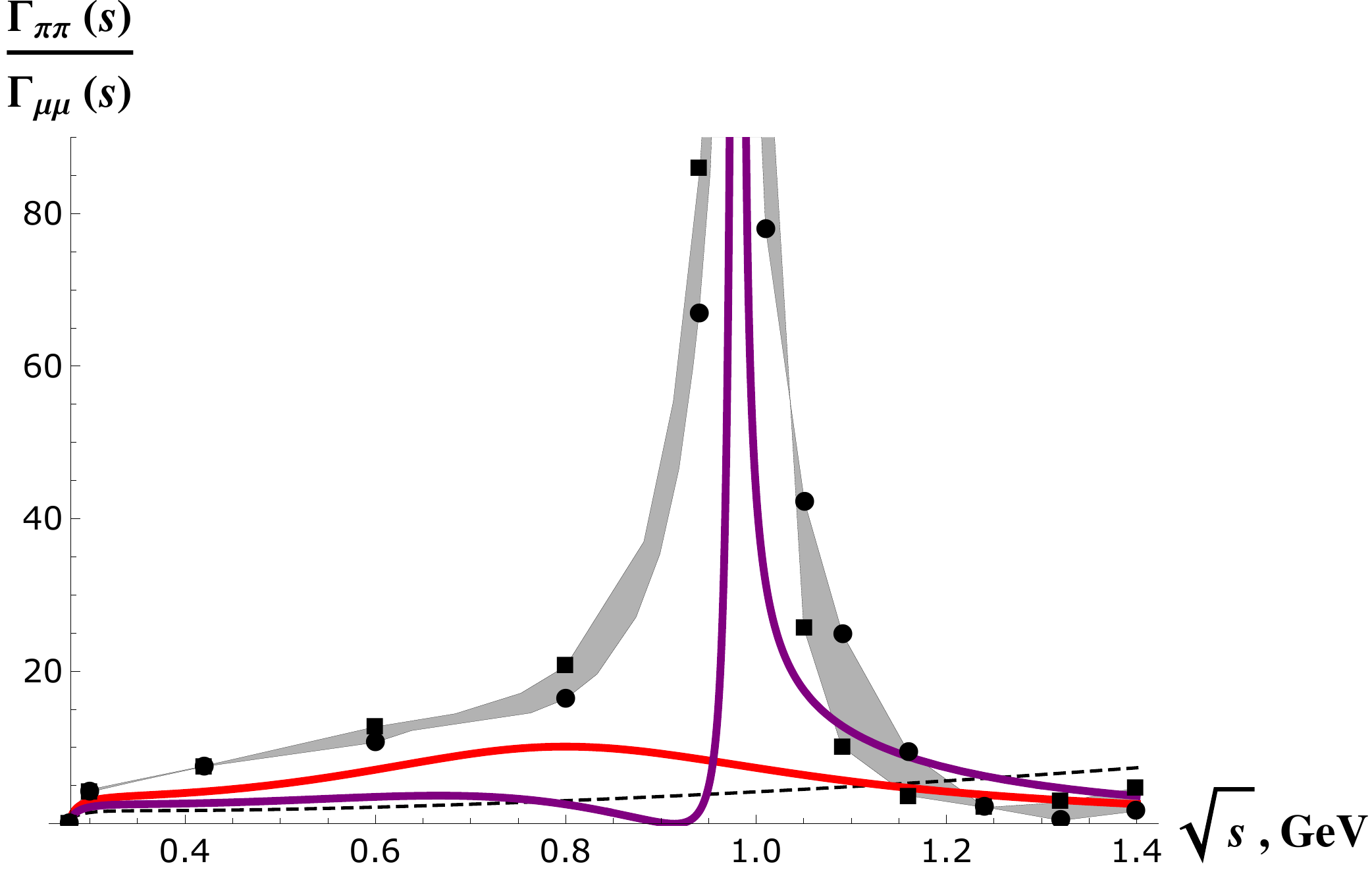}}
\ec
\caption{\label{fig:decayRate} Branching ratio as a function of mass $m_S=\sqrt{s}$: ChPT (dashed black), one-channel resonance approximation with parameters from~\cite{Raby:1988qf} (red), data points and black curves are taken from~\cite{Donoghue:1990xh}.}
\end{figure}

Lastly, we present the results for the decay rate in Fig.~\ref{fig:decayRate}. For both, one- and modified one-channel approximations (\ref{eq:1chD}) and (\ref{eq:1chModFF}) we use $d_F=0.09$ from~\cite{Donoghue:1990xh}. A somewhat different value (also discussed in~\cite{Donoghue:1990xh}), obtained in the limit of heavy s-quark, $d_F=2/29\approx 0.068$, changes the result only slightly. As is also evidenced from plots for the form factors (Figs.~\ref{fig:1chCut1chMod}, \ref{fig:thetaFF2} and \ref{fig:DeltaFF}) the modified one-channel approximation with an additional zero (\ref{eq:addZero}) characterized by $\sqrt{s_{\t}}=1.3$~GeV and $\sqrt s_\G = 1$~GeV reproduces numerical results of~\cite{Donoghue:1990xh} (Fig.~\ref{fig:1chMod}). It is clear that all methods (except the two-channel resonance approximation) produce results significantly different from the ChPT just above $4m_\pi^2$. We take that as an indication that ChPT results should be modified at rather low energies. Similar situation occurs in the case of $\eta \to \pi^0 \pi^+ \pi^-$ decay, where the final state interaction effects are proven to be large~\cite{Roiesnel:1980gd}.

\section{Discussion}
\label{sec:discussion}

\subsection{A rough error estimate}

Unarguably the two-channel numerical solution makes use of the experimental
data the most, therefore, it should produce the most reliable results. At the
same time it is important to quantify the uncertainties and to figure out the
domain of validity of the result.
Even though examples of Section~\ref{sec:resonance} (see Fig.~\ref{fig:2chRes_analysis}) show that form factors may appreciably depend on the parameters in (\ref{2ch_ff}) we will not discuss these corrections, referring the reader to~\cite{Donoghue:1990xh} claiming that variations of the parameters (\ref{2ch_ff}) does not influence the decay rate significantly. Similarly, the dependence of the result on different inputs for the phase shifts and different interpolations of the data is not discussed, since it was addressed in~\cite{Donoghue:1990xh} (see Fig.~\ref{fig:decayRate}).

Instead we discuss another source of uncertainty, which plagues all the discussed methods. It is our ignorance about the UV dynamics, for we do not know the S-matrix above $\sqrt{s}=\Lambda_\text{dat}$. The importance of scattering data above that scale is clear, for none of the methods (except the modified one-channel approximation taken at face value) produce the correct asymptotic of the form factors. This is an indication that these methods cannot present the complete picture. The role of high energy dynamics is clearly important.

From a more technical perspective the knowledge of the high energy behavior of the scattering amplitudes is crucial in all the dispersion methods, involving integrals to infinity. Qualitatively if the UV dynamics kicks in at a relatively high energy scale it should not affect the form factors in the IR, it is decoupled. However, quantitatively we need to have at least a rough error estimate.

To do that instead of solving the system of two coupled channels with a different asymptotic for scattering phases it is easier to consider the approximation to the solution provided by the modified one-channel approximation (\ref{eq:1chModFF}) and (\ref{eq:addZero}) which is amenable to analytic treatment. The scattering phase is known up to $s=\Lambda^2_\text{dat}$. We assumed that at infinity it asymptotes to $2\pi$. The ratio between the Omn\`es factor obtained with the help of so defined phase (\ref{Omnes-solution-1ch}) and the physical one is given by
\be
\f{\Omega_\pi(s)}{\Omega^\text{phys}_\pi (s)} = \exp \l [ \f {s} {\pi} \int _{\Lambda_\text{\blue dat}^2}^{\infty} \f{d s'}{s'} \, \f{\d_\pi(s') - \d^\text{phys}_\pi(s')} {s' - s - i\eps} \r].
\ee

We are focusing only on situations when 
the asymptotic behavior is different. In this case it is easy to estimate the correction at small $s$ has the form
\be
\Omega^\text{phys}_\pi(s) \approx \Omega_\pi(s) \l ( 1+ c \f {s} {\Lambda_\text{diff}^2} \r ),
\label{eq:errorEstimate}
\ee
where $\Lambda_\text{diff}$ is the scale where the value of $\d_\pi(s)$ and $\d^\text{phys}_\pi(s)$ differ by  $\pi$ (e.g. where a new resonance comes into play) and $c$ is an order one constant.

\subsection{Domain of validity\label{sec:domain}}

Armed with (\ref{eq:errorEstimate}) we can estimate how reliable the results are at a specific scale. And the benchmark is set by other corrections compared to (\ref{eq:errorEstimate}). The derivation of Section~\ref{sec:setup} is performed at the leading order in the heavy quark mass $m_c^{-2}$ and the strong coupling $\a_s$. The next to leading corrections can be found correspondingly in~\cite{Shifman:1978zn} and~\cite{Barbieri:1988ct}. The relative correction due to QCD loops is estimated to be  $5 - 10\%$~\cite{Barbieri:1988ct}.
At the same time we see from Fig.~\ref{fig:decayRate} that the discrepancy in the decay rate between different solutions obtained in~\cite{Donoghue:1990xh} does not exceed $30\%$ up to $1~$GeV. Therefore, the relative error of $10-15\%$ for form factors seems to be a good starting point\footnote{As discussed above in the interval $16 m_{\pi}^{2}<s < \Lambda_{\dat} ^2$ the restriction of the $S$-matrix to two channels is only approximately unitary. Mixing with multi-pion states leads to about $6\%$ corrections according to~\cite{Yndurain:2003vk}.}. And the scale $\sim \Lambda_\text{diff}/4$ is where the naively estimated corrections due to $UV$ phase shifts (\ref{eq:errorEstimate}) exceed this benchmark.

\begin{figure}[t]
\bc
\subfloat[\label{fig:Delta-1ch-Error} $\D_\pi(s)$ form factor.]{\includegraphics[width=0.3\textwidth]{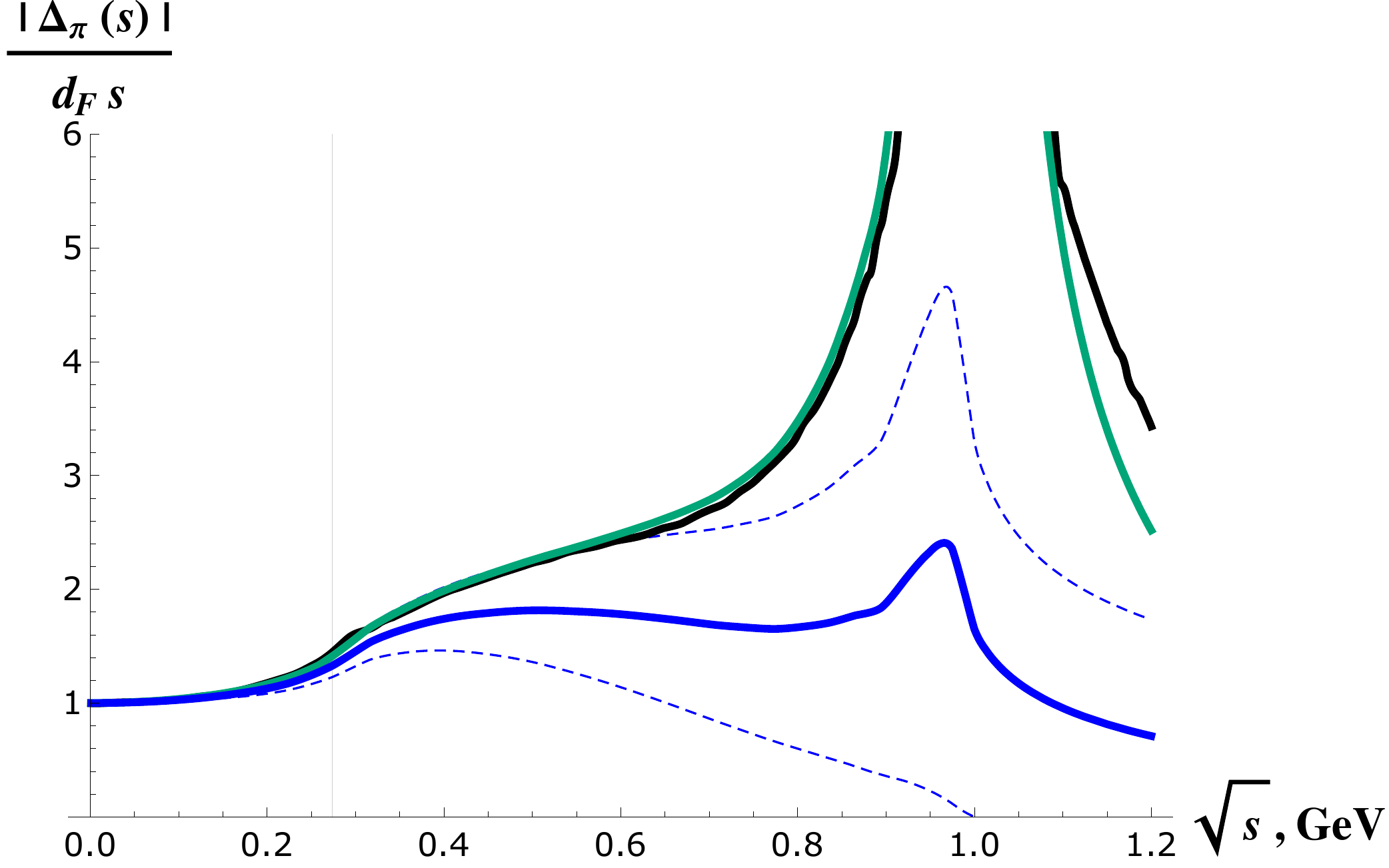}}
\hfill
\subfloat[\label{fig:Gamma-1ch-Error} $\G_\pi(s)$ form factor.]{\includegraphics[width=0.3\textwidth]{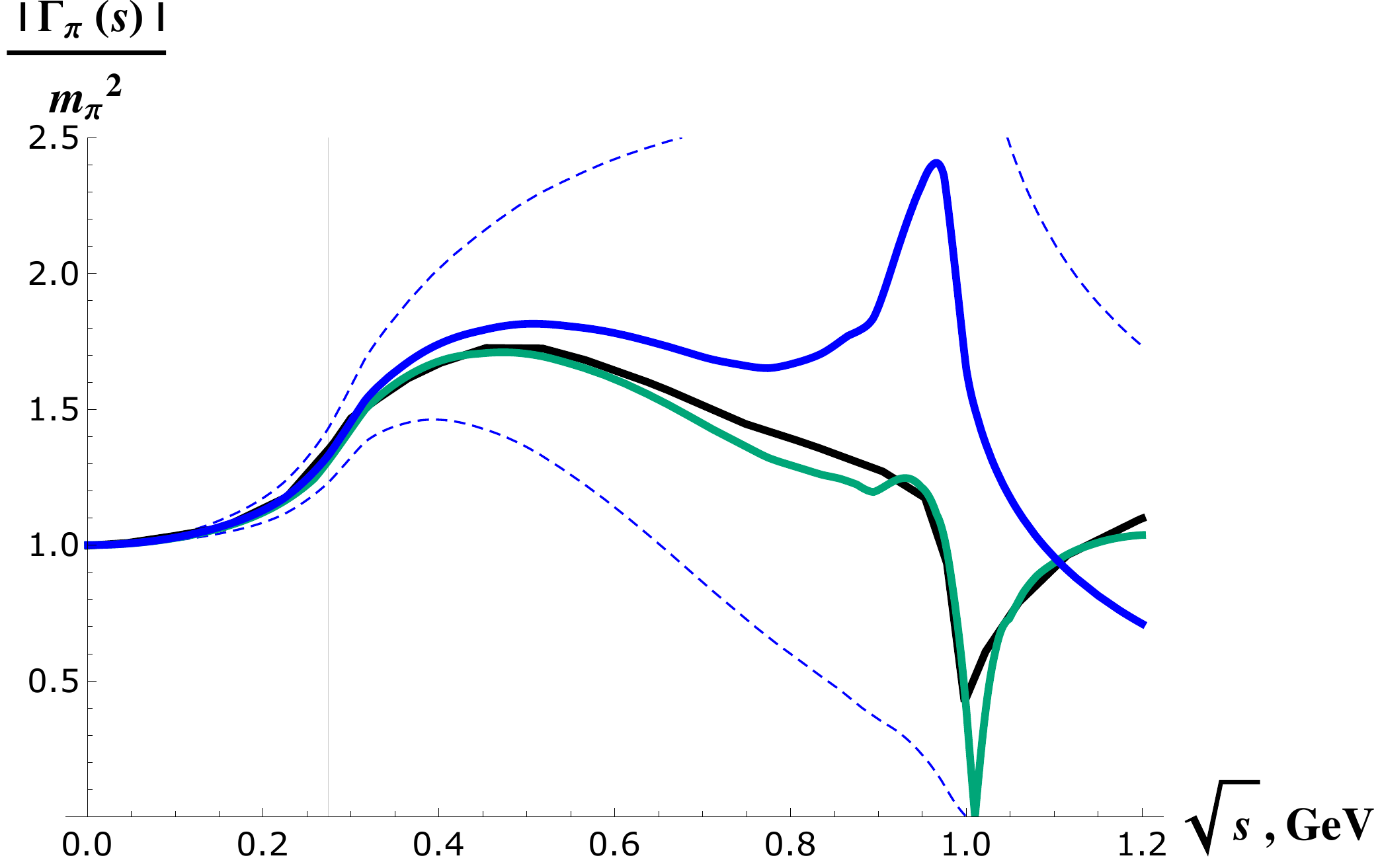}}
\hfill
\subfloat[\label{fig:Width-1ch-Error} Branching ratio.]{\includegraphics[width=0.3\textwidth]{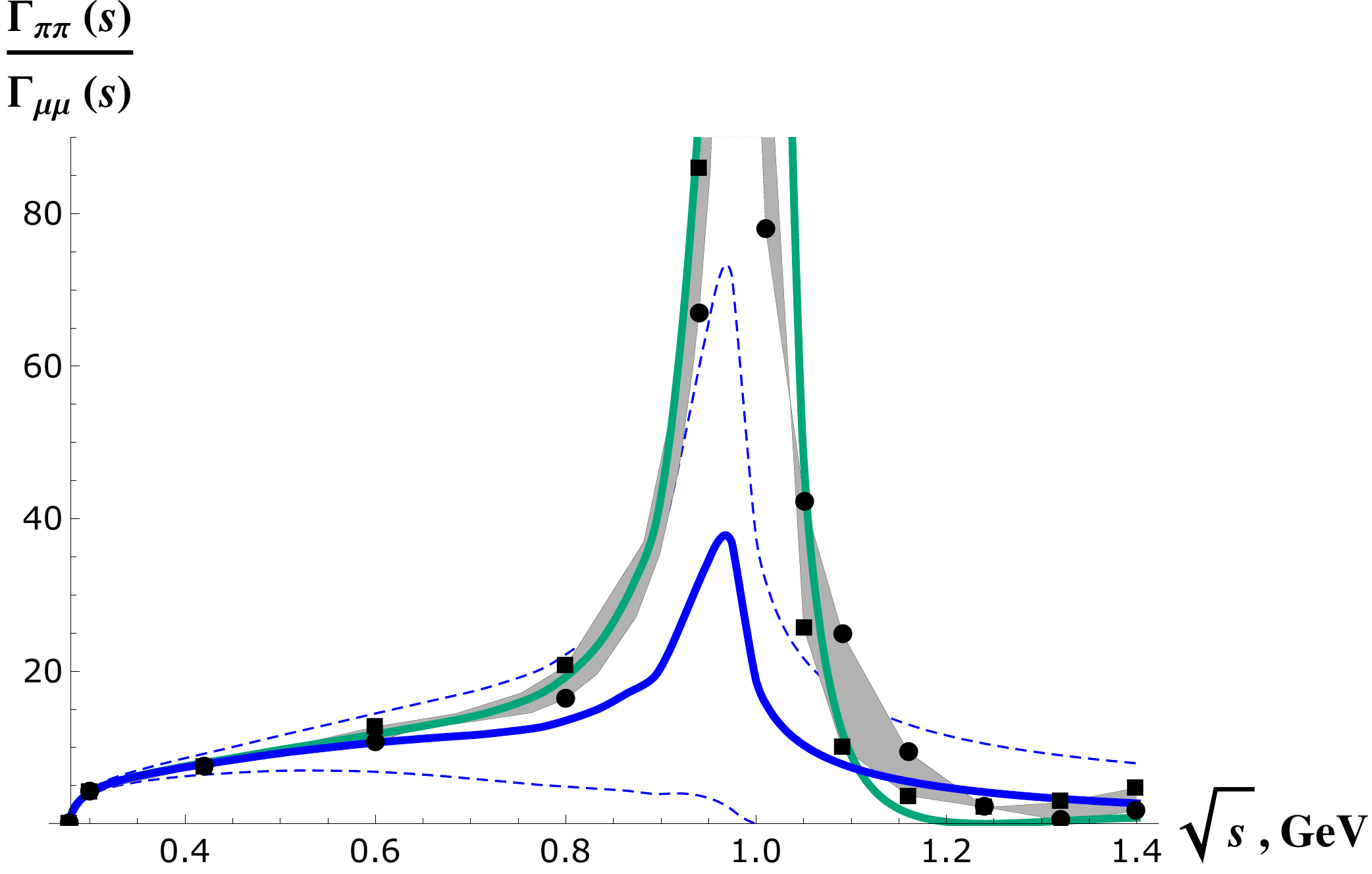}}
\ec
\caption{\label{fig:1ch-Error} \textbf{Error estimates:} results for $\Gamma_\pi$ (a), $\Delta_\pi$ (b), and the total decay width $\Gamma_{\pi\pi}$ (c) in the one-channel approximation (blue) with a relative error (\ref{eq:errorEstimate}) shown in blue dashed.
  Two-channel numerical results are shown in black: panels (a) and (b) from~\cite{Ananthanarayan:2004xy} and panel (c) from~\cite{Donoghue:1990xh}.
  Modified one-channel approximation with an additional zero is shown in green.}
\end{figure}

As an example we see from Fig.~\ref{fig:Delta-1ch-Error} that the behavior of the $\D_\pi(s)$ form factor is perfectly consistent with (\ref{eq:errorEstimate}) with $c=1$ and $\Lambda_\text{diff}^2 = 4m_K^2$. The reason for that is the occurrence of a rather narrow resonance $f_0(980)$, leading to a sharp increase in the scattering phase around kaon threshold. At the same time Fig.~\ref{fig:Gamma-1ch-Error} shows that even though the actual two-channel result lies within corrections dictated by (\ref{eq:errorEstimate}), the latter ones are obviously an overestimate. And the scale where one- and two-channel results deviate from each other is larger than is expected naively, which effectively means that in this case $c<1$. That in turn leads to a better agreement between the results for the decay rate.

To estimate how wide the domain of validity is for two-channel numerical analysis of~\cite{Donoghue:1990xh} (equivalently the modified one-channel approximation) it is necessarily to know whether similar to $f_0(980)$ resonances above $\Lambda_\text{dat}$ play a significant role for scattering phases. While $f_0(1370)$ is rather wide and we may not see its effect, there are claims that $f_0(1500)$ is visible in the scattering data~\cite{Kaminski:1998ns}. At the same time it is unknown whether the resonances $f_0(1710)$, $f_0(2110)$, $f_0(2200)$ and $f_0(2330)$, listed in~\cite{pdg_data_res}, affect the scattering phase. As a result, if indeed $f_0(1500)$ affects the scattering phase, the cutoff $\Lambda_\text{diff}$ would be given by $1.5$~GeV. On the other hand, if none of those resonances change the behavior of the scattering phase
the cutoff will be at least $2.5$~GeV. 

\begin{figure}[!t]
\bc
\subfloat[\label{fig:Width-1chMod-15-Error} $\Lambda_\text{diff}=1.5$~GeV.]{\includegraphics[width=0.45\textwidth]{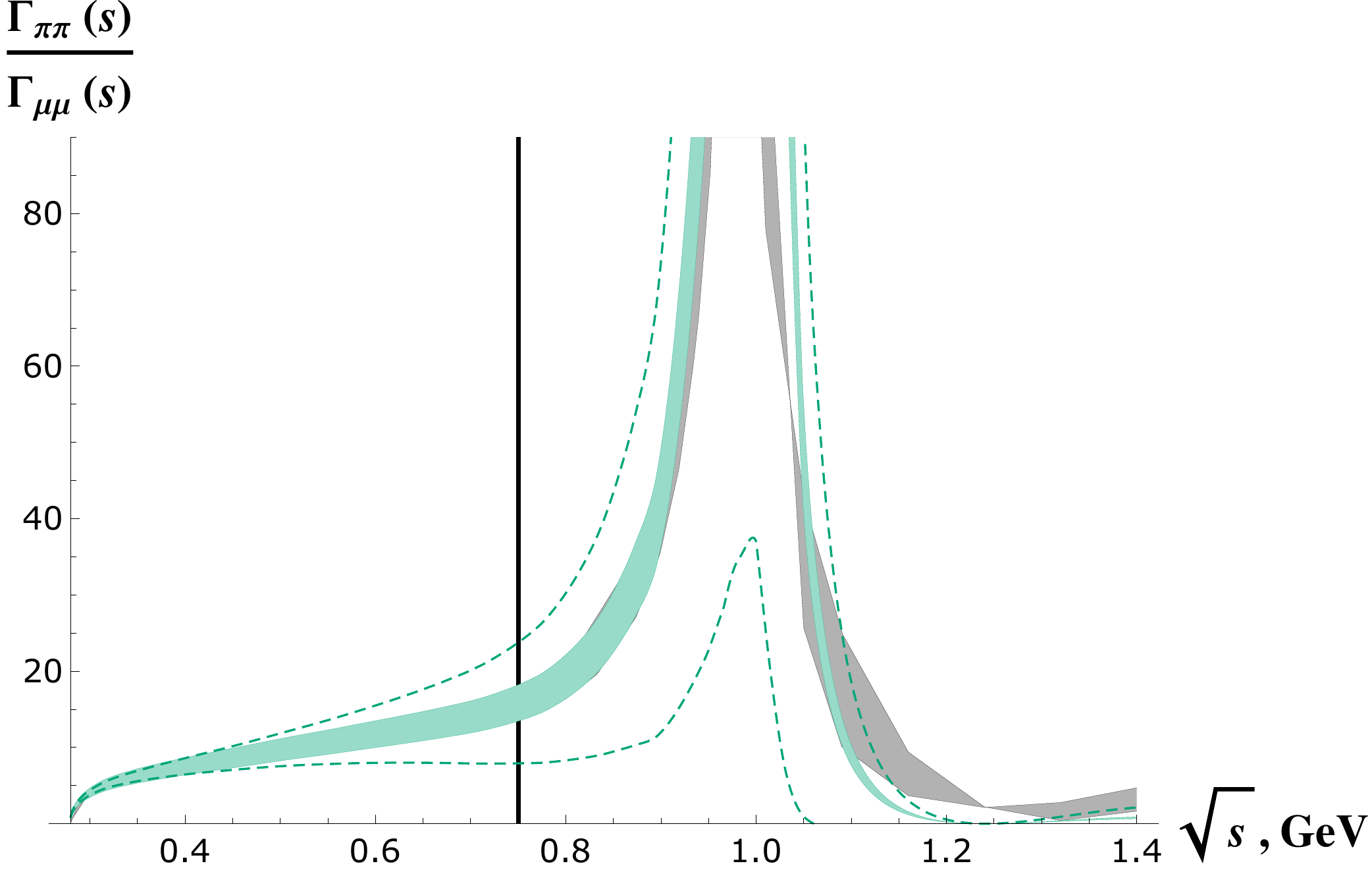}}
\hfill
\subfloat[\label{fig:Width-1chMod-25-Error} $\Lambda_\text{diff}=2.5$~GeV]{\includegraphics[width=0.45\textwidth]{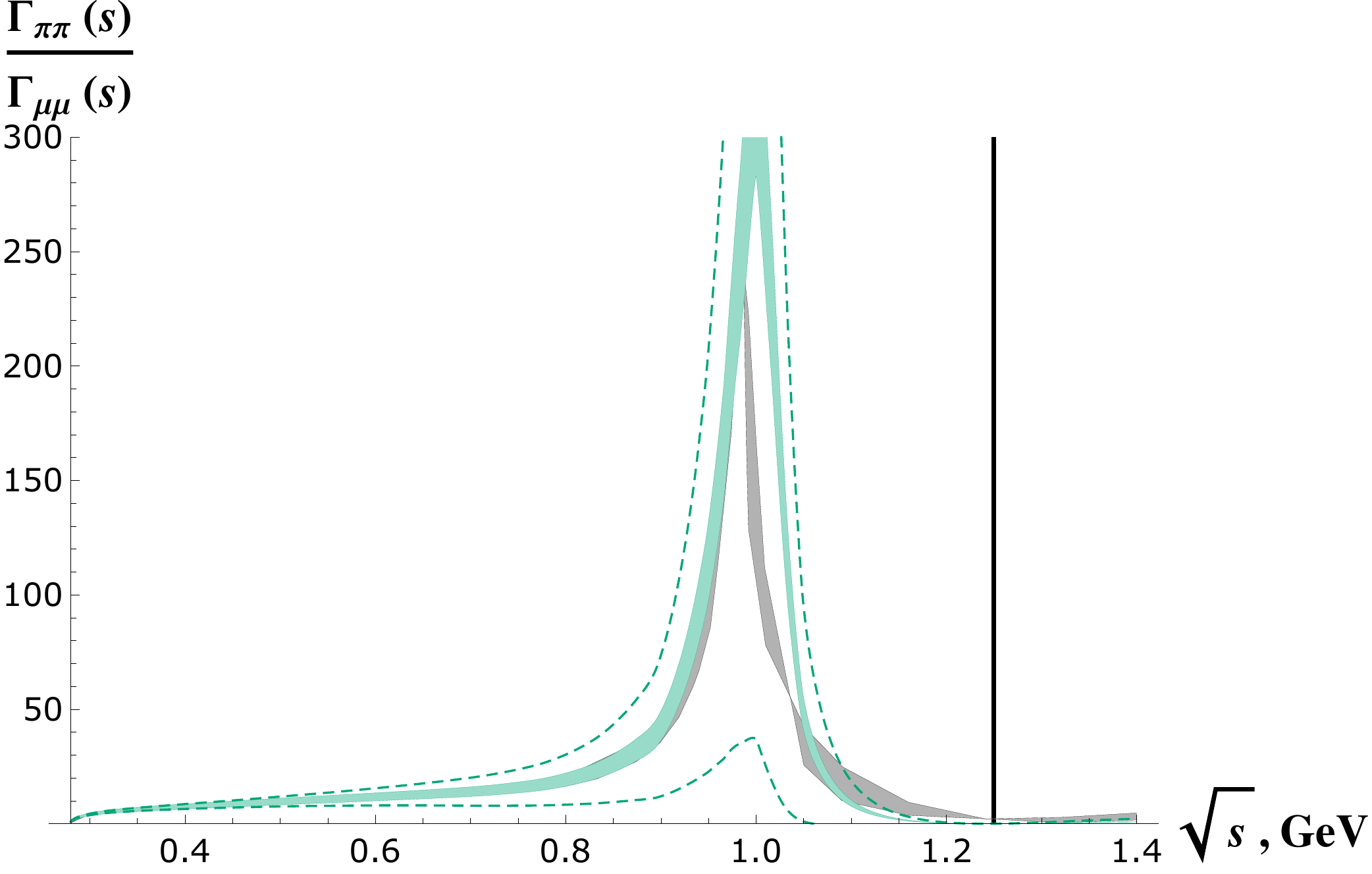}}
\ec
\caption{\label{fig:1chMod-Error} Branching ratio from~\cite{Donoghue:1990xh}
  (two black curves corresponding to different parametrizations) and the
  modified one-channel approximation (green) with parameters chosen so as to
  approximate these results: $d_F = 0.09$, $s_\t = 1.2~$GeV$^2$, $s_\G=1~$GeV$^2$. Shaded green regions indicate $30\%$ error-bars. Dashed lines correspond to naively corrected results ($1\pm s/\Lambda_\text{diff}^2$) according to~(\ref{eq:errorEstimate}): vertical lines show where the leading linear in $s$ correction becomes 100\%.}
\end{figure}

We plot the decay rate as a function of the scalar mass $m_S$ with the corresponding corrections in Fig.~\ref{fig:1chMod-Error}. We note that the error estimate (\ref{eq:errorEstimate}) should be consider as an upper bound, for as in Figs.~\ref{fig:Gamma-1ch-Error}, \ref{fig:Width-1ch-Error} the actual corrections, with the dynamics of all channels taken into account, can be noticeable reduced.
The reason is that changing the asymptotic of the scattering phase by $\pi$ necessitates using a polynomial of one degree higher (introducing an additional zero) to preserve the correct asymptotic behavior of the form factor at infinity, unless it was incorrect in the first place. The relative correction results in the two effects (partially) cancelling each other
\be
\l ( 1 + c \f {s}{\Lambda_\text{diff}^2} \r ) \l ( 1 - \f {s}{s_0}\r ) = 1 +c \f {s}{\Lambda_\text{diff}^2} - \f {s}{s_0}.
\ee
We see that the error depends on where the position $s_0 \geq \Lambda_\text{diff}^2$ of the additional zero is. If it is very close to the cutoff $s_0 \approx \Lambda_\text{diff}^2$ -- what happens for $\G_\pi^\text{mod}(s)$ -- the error is significantly reduced. While if it lies far in the UV or the behavior of the form factor was incorrect before modifying the phase, hence, there is no need to use a different polynomial with an extra zero (this is precisely the case for $\D_\pi^\text{1ch}(s)$), the error is given by (\ref{eq:errorEstimate}).

\subsection{Experimental data for $\psi '$ and $\Upsilon'$ decays}

\begin{figure}[h]
\bc
\includegraphics[width=10cm]{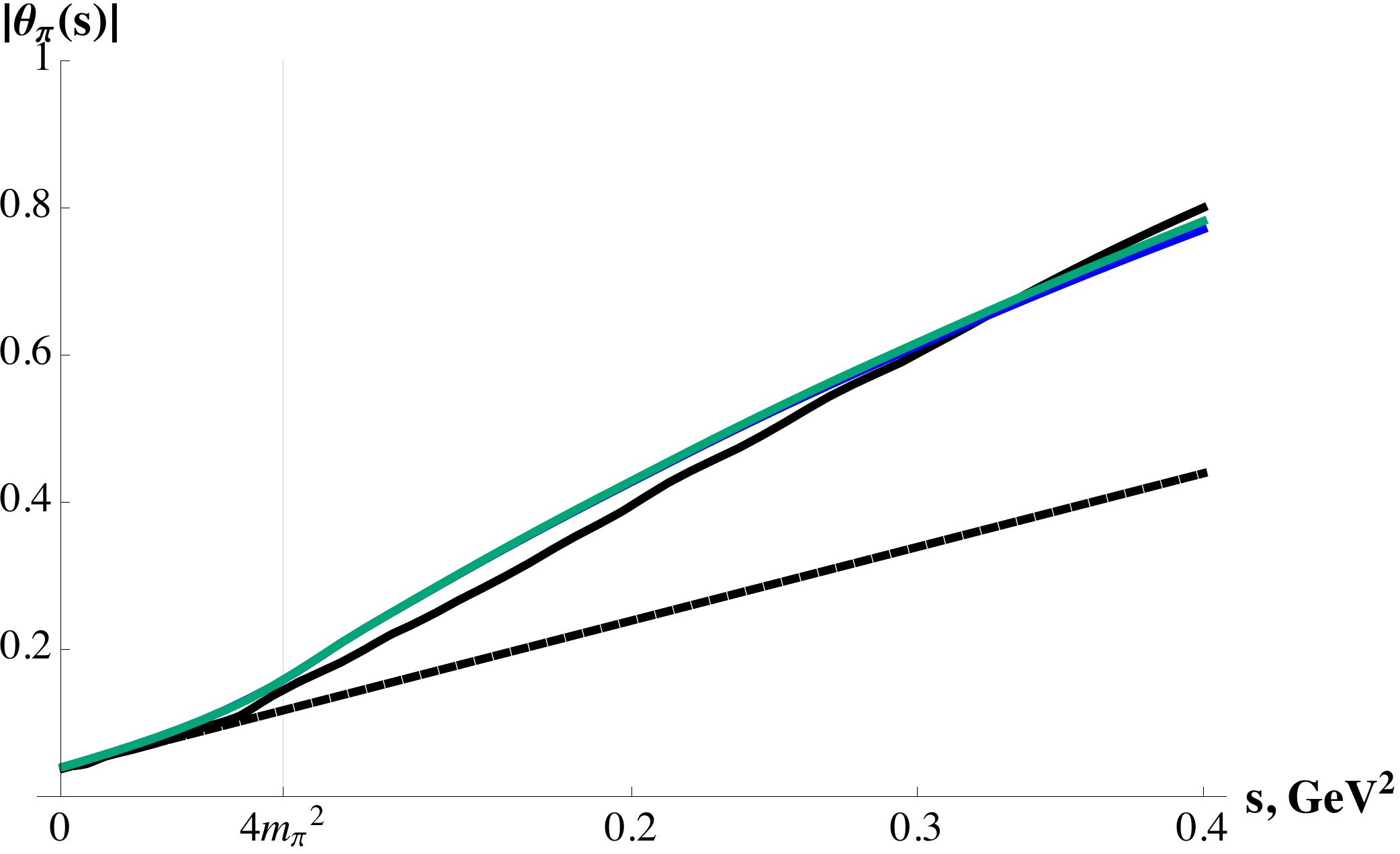}
\caption{Approximately linear behavior of $\t_\pi(s)$: One-channel solution (blue), modified one-channel (green), two-channel numeric solution (solid black), ChPT (black dashed).  \label{theta_ChPT} }
\ec
\end{figure}

Ref.~\cite{Voloshin:1985tc} presents arguments in favor of extrapolating the ChPT result up to energies $\sqrt{s}=0.6~\text{GeV}$. It assumes that the decay rate $\G_{\pi\pi}$ is dominated by the form factor $\theta_\pi(s)$. At the same time, according to the data from $\Upsilon'\to \Upsilon \pi \pi$~\cite{Albrecht:1983we} and $\psi' \to \psi \pi \pi$~\cite{Bai:1999mj, Himel:1979dj} decays, the form factor $\theta_\pi(s)$ behaves linearly in the relevant interval of energies~\cite{Voloshin:1975yb,Voloshin:2006ce,Voloshin:2007dx}. Thus, it is tempting to conclude
that there is no deviation from ChPT prediction at these energies. However (see Fig.~\ref{theta_ChPT}), the behavior of the form factor $\t_\pi(s)$ obtained via dispersion relations also close to linear.
It would be helpful to find the slope of the form factor in the interval $2m_\pi \leq \sqrt{s} \lesssim 0.6~\text{GeV}$ using the data for mentioned decays.
That could potentially resolve the issue and will be considered elsewhere.

\section{Conclusion\label{sec_conclusion}}

In this work we have revisited a problem of computing hadronic decay width of a Higgs-like scalar with mass of order $m_S \sim 1~\text{GeV}$.
At these energies decays $S\to \pi\pi$ or $S\to \bar K K$ cannot be adequately described by ChPT.
It was previously demonstrated~\cite{Donoghue:1990xh} that non-perturbative methods, reconstructing the form factors from the data on meson scattering, yield the results that are indeed different from the leading order ChPT calculations~\cite{Voloshin:1985tc}, significantly enhancing the decay rate starting from several hundreds of MeV.
The tradeoff is that the data on meson scattering, used as an input, should be provided for arbitrarily high energies, which is of course not possible.
In practice having the data only up to a certain energy produces an approximation to the exact function.
The assessment of the uncertainties of the estimate~\cite{Donoghue:1990xh} have never been performed before.

The results of this work are summarized as follows.
We reconsidered the non-perturbative evaluation of the hadronic decay width paying special attention to the uncertainties of the method.
Even though above $1$~GeV there are at least two relevant channels ($\pi\pi$ and $\bar KK$), in this paper we used the (modified) one-channel approximation to reproduce the results of the two-channel numerical analysis of~\cite{Donoghue:1990xh}.
This one-channel approximation works well because the elasticity parameter, controlling the coupling between the two channels, differs from $1$ only in a narrow region above the kaon threshold.
That effectively renders the two channels independent outside of this narrow region.
The approximate one-channel solution is advantageous as it is known
analytically for any scattering phase.
The
corresponding parameters reproducing~\cite{Donoghue:1990xh} are $d_F = 0.09$, $s_\t = 1.2~$GeV$^2$, $s_\G=1~$GeV$^2$.
Therefore, it is straightforward to estimate corrections coming from different UV behaviors of the scattering phase.
We analyzed different scenarios and our main result is presented in
Fig.~\ref{fig:1chMod-Error}, showing the upper bound for the error.

It seems unarguable that the leading ChPT result proves to be inadequate in describing the process (even in a close vicinity of $\pi \pi$ threshold).
The reason is (see also~\cite{Donoghue:1990xh}) the presence of a rather broad $f_0(980)$ resonance.
The disproportionate enhancement of the form factor $\D_\pi(s)$ compared to $\G_\pi(s)$ and $\t_\pi(s)$ indicates an augmented relative significance of the s-quark effects for the process at hand.
If resonances above $1.5$~GeV couple to the channel in question and change the scattering phase, the behavior of the form factors and, hence, of the decay rate will be affected significantly around those resonances (as is evidenced in Appendix~\ref{sec:modelPhase}), but will be intact for sufficiently small energies.

\section*{Acknowledgements}

We are extremely grateful to Profs G.~Colangelo, M.~Shaposhnikov and M.~Voloshin for discussions and valuable comments. However, all the mistakes in the present paper are solely our responsibility. The work of AM is supported by the ERC-AdG-2015 grant 694896 and the Swiss National Science Foundation Ambizione grant.

\appendices

\section{Extrapolating the leading ChPT result \label{sec:ChPTextrapolation}}

The estimate of the scale where the leading ChPT result is based on the paper~\cite{Shifman:1980iq} and done as follows. It is assumed that the decay rate is dominated by $\t(s)$ and is thus related to the spectral density
\begin{equation}
\rho (q^{2}) = (2\pi)^{4} \sum _{n} | \la 0 | \t _{ \m}^{\m} (0) | n \ra | ^{ 2} \d ^{4} (q - q _{n}),
\label{spec-dens}
\end{equation}
evaluated at $q^{2}=m_{S}^{2}$. The spectral density is given by the imaginary part of the two-point correlation function of the energy momentum tensor (\ref{conf-an}), which in turn in the chiral limit is expressed in terms of the gluon field strength
namely,
\begin{equation}
\rho (q^{2}) = 2 \Im \Pi (q^{2}) = 2 i \int dx \,  e ^{ i q x} \la \mathrm T \, \f {\beta (\a)} {4 \a} G ^{ 2} (x) \, \f {\beta (\a)} {4 \a} G ^{ 2} (0) \ra.
\end{equation}

At low energies spectral density (\ref{spec-dens}) is saturated by Goldstone bosons and can be computed taking into account (\ref{low-energy-theta-ff})
\begin{equation}
\rho(s) = \f {N_{\ell}^{2}-1} {16 \pi} s^{2} + O (s ^{3}),
\label{spec-dens-low}
\end{equation}
with $N_{l}$ being the number of light quarks. On the other hand it follows (see~\cite{Shifman:1980iq}) that the following sum rule is satisfied
\begin{equation}
\int \f{ds}{s} \l [ \rho (s) - \rho _{\text{pert}} (s) \r ] = b \a \, \la G ^{2} (0) \ra,
\label{sum-rule}
\end{equation}
where $\rho _{\text{pert}} (s)$ is perturbative contribution, whose counterpart on the r.h.s.
is implicitly present in the form of the renormalization of $\la G ^{2} (0) \ra$ needed to make it finite.
It is expected that for large enough energies $\rho(s)$ is accurately represented by $\rho_{\text{pert}}(s)$. Therefore, the integral (\ref{sum-rule}) is saturated at low energies.
Thus doing the integral with the LO approximation (\ref{spec-dens-low}) and taking into account the value for $\la G ^{2} (0) \ra$ from~\cite{Shifman:1978by,Novikov:1979va} allows one to estimate the cutoff where the true value of $\rho(s)$ substantially deviates from the LO (\ref{spec-dens-low}), which happens to be around 1.4 GeV.

The reasoning presented above is somewhat reminiscent of that for the fine tuning problem and certainly gives an upper bound for when the LO approximation breaks down. However, it may happen earlier. Higher order in momentum corrections may kick in to bring the scale, where the approximation cannot be trusted anymore, down.

\section{Modelling phase \label{sec:modelPhase}}

\begin{figure}[h!]
\bc
\subfloat[\label{fig:phaseModel1} Resonance at $\Lambda_\text{diff}=2.5$~GeV.]{\includegraphics[width=0.3\textwidth]{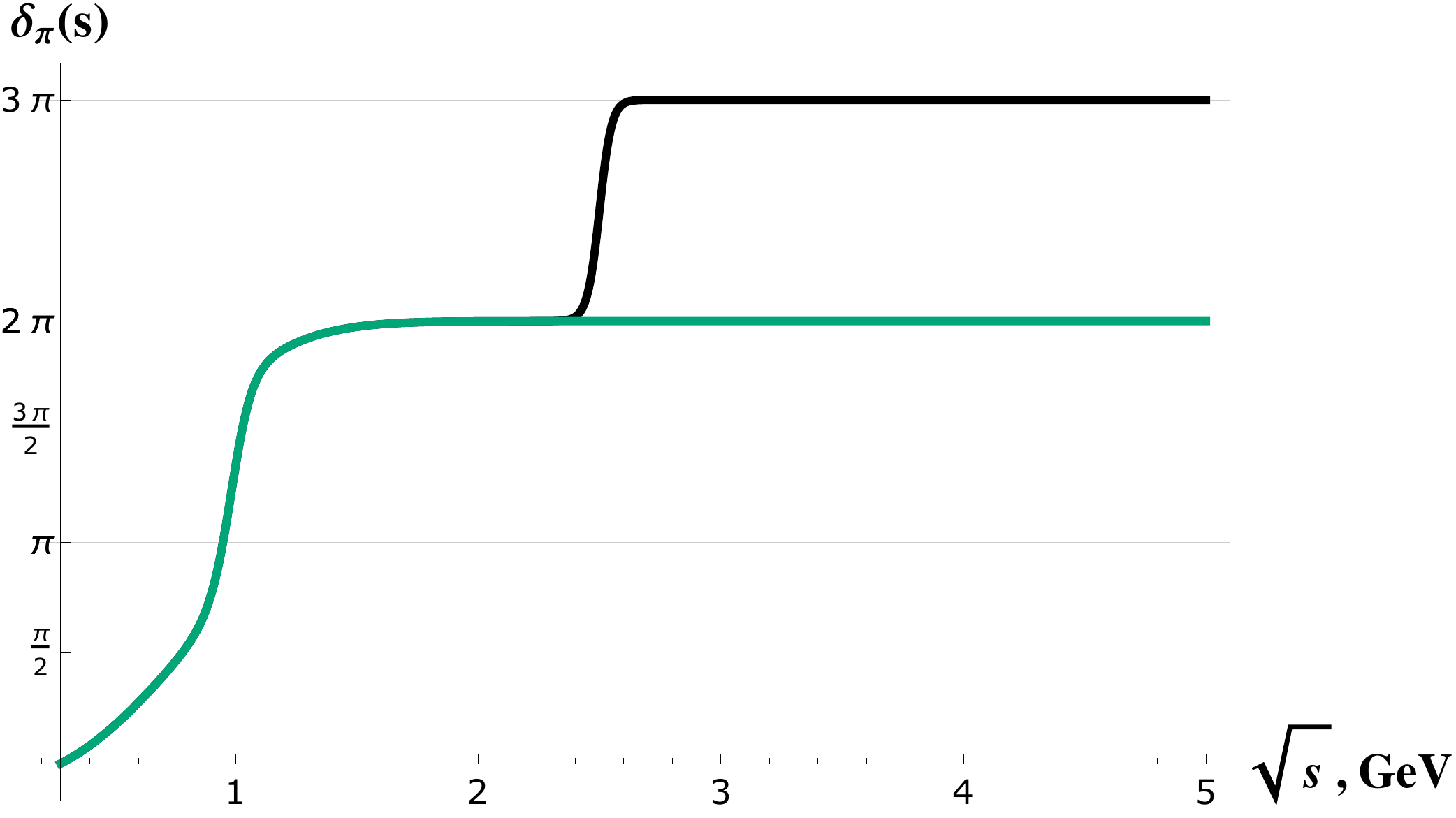}}
\hfill
\subfloat[\label{fig:phaseModel2} Resonance at $\Lambda_\text{diff}=1.7$~GeV.]{\includegraphics[width=0.3\textwidth]{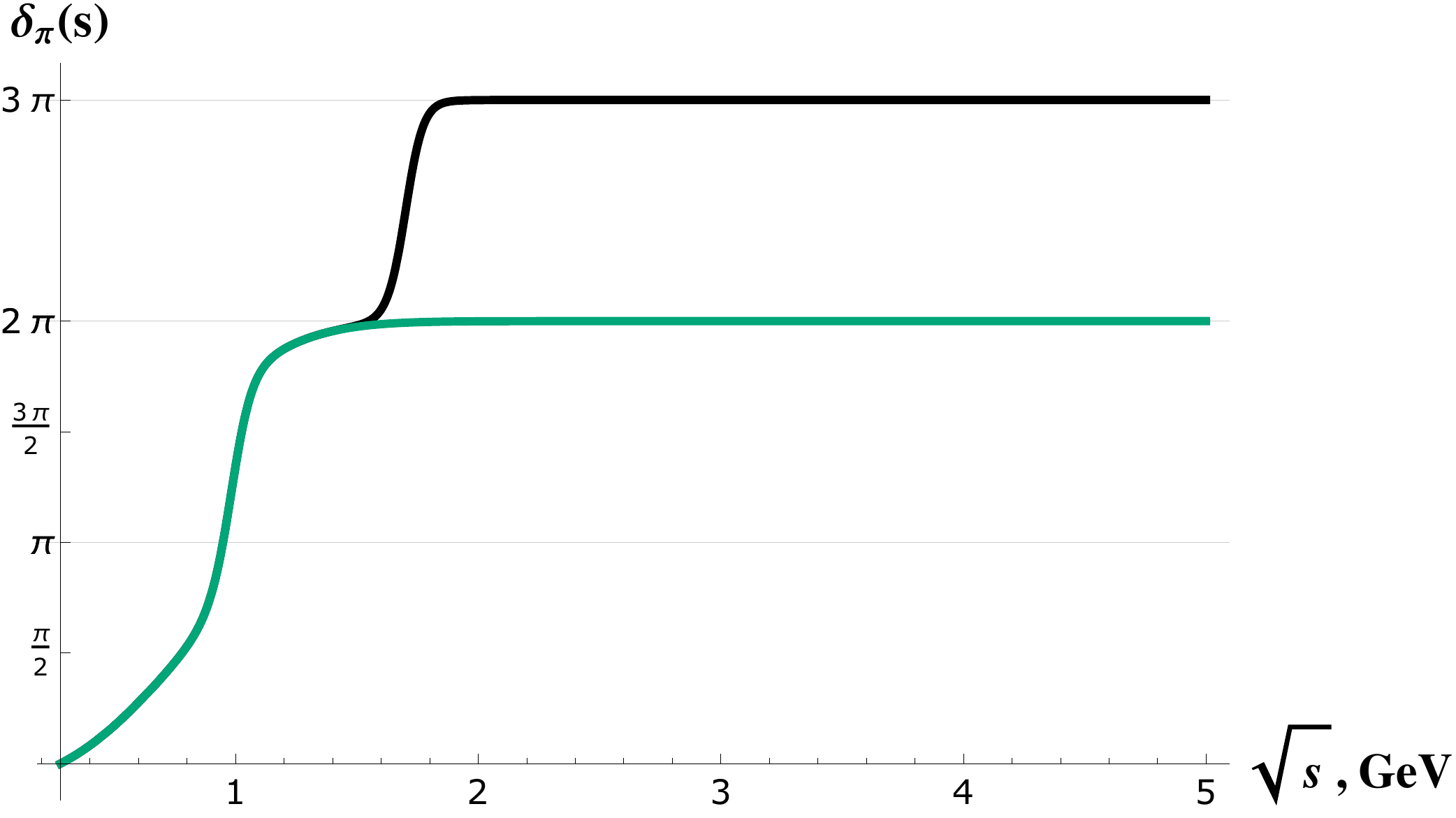}}
\hfill
\subfloat[\label{fig:phaseModel3} Resonances at  $1.7$~GeV and $1.9$~GeV.]{\includegraphics[width=0.3\textwidth]{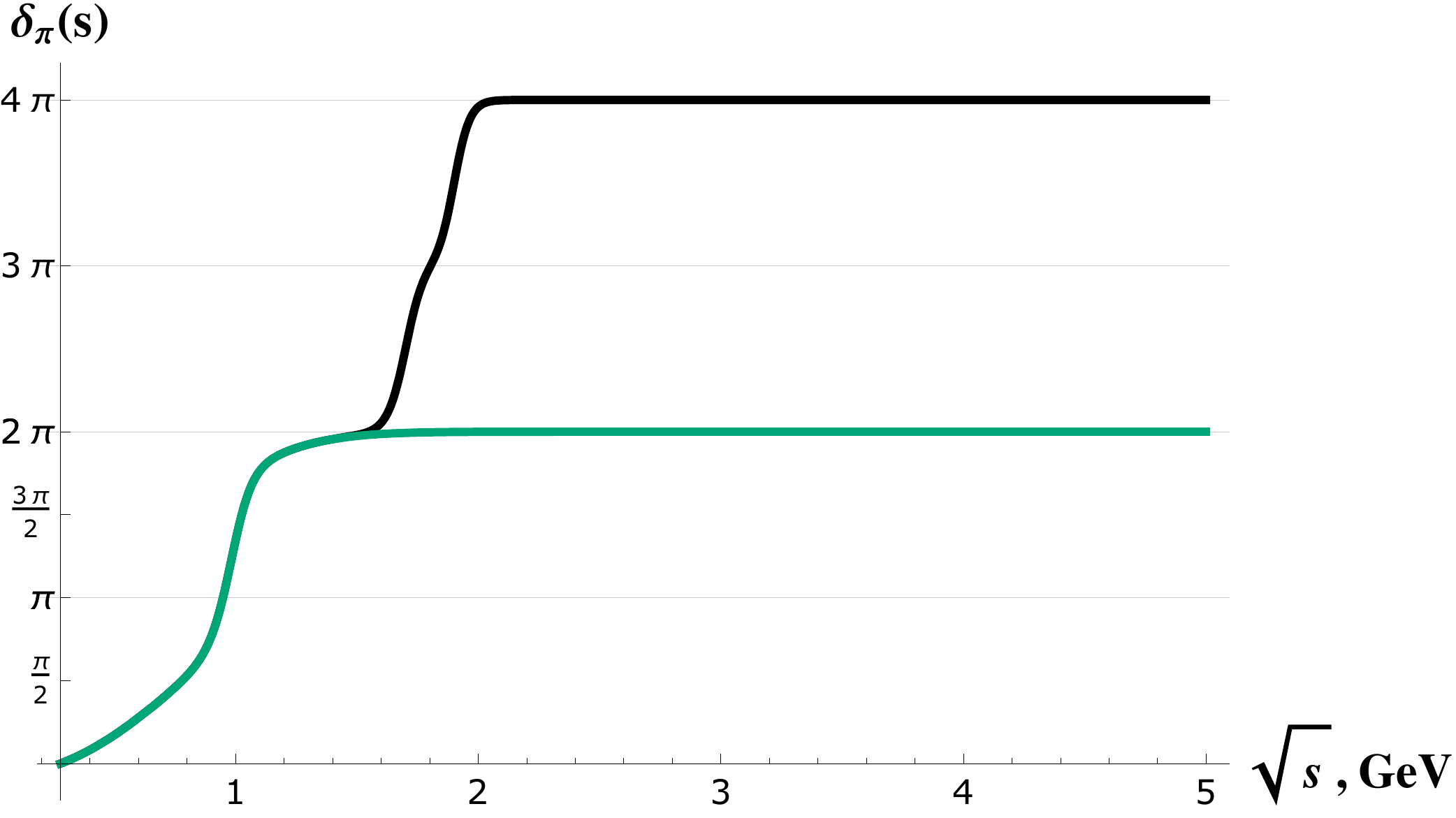}}
\ec
\caption{\label{fig:phaseModels} Scattering phase behavior modelled by several resonances: "experimental" data (green), actual phase (black).}
\end{figure}

In this section we demonstrate how different UV behaviors of the scattering phase affect form factors. For that we consider the following hypothetical situation. We assume that the scattering phase is known up to $1.5$~GeV, and it is extrapolated to $2\pi$ above\footnote{The phase is chosen to qualitatively reproduce the real world behavior.}. The actual behavior of the physical phase at the same time is modelled by  three different scenarios depicted in Fig.~\ref{fig:phaseModels}.

\begin{figure}[h]
\bc
\subfloat[\label{fig:farRes1} $\Lambda_\text{diff}=2.5$~GeV.]{\includegraphics[width=0.45\textwidth]{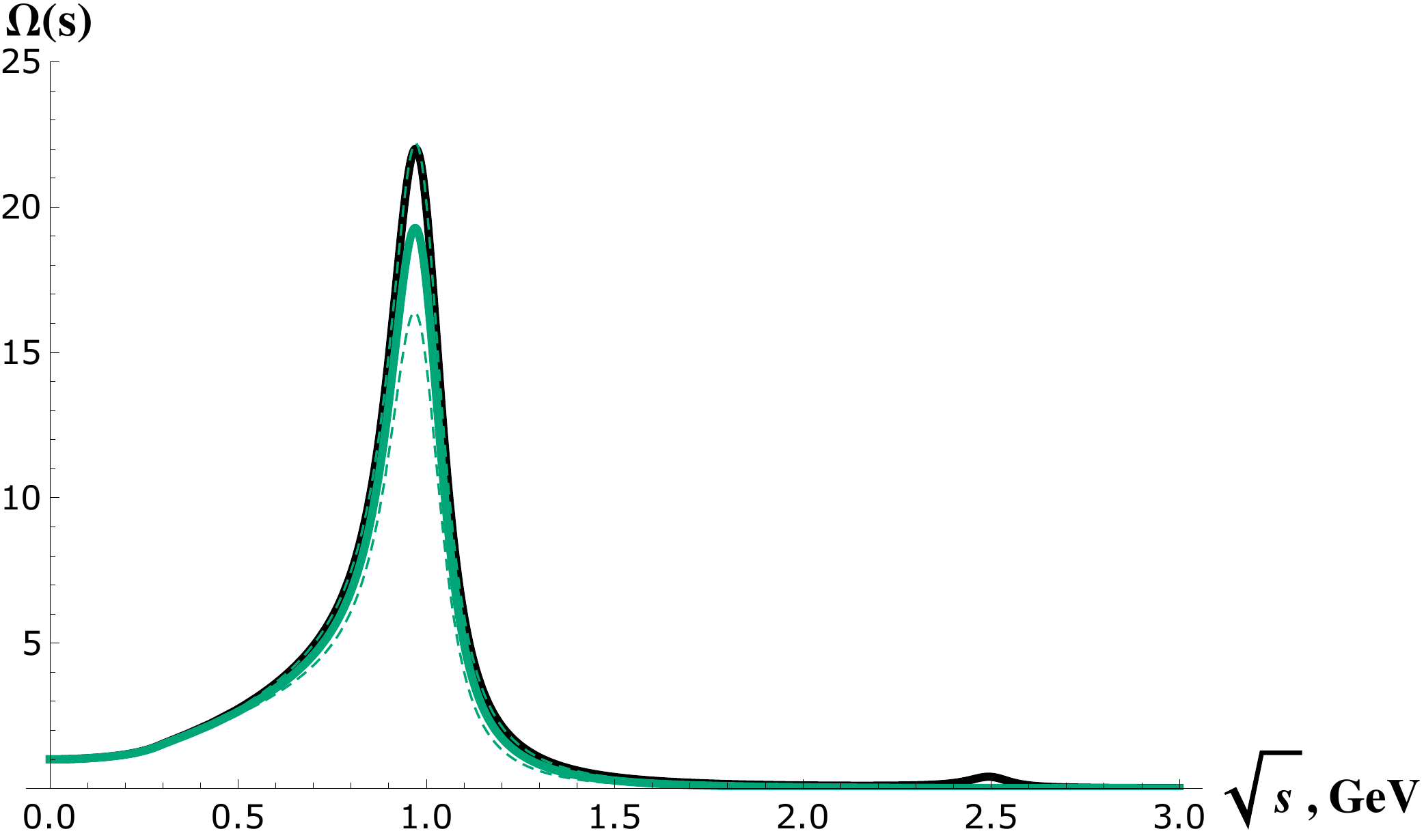}}
\hfill
\subfloat[\label{fig:farRes2} $\Lambda_\text{diff}=2.5$~GeV]{\includegraphics[width=0.45\textwidth]{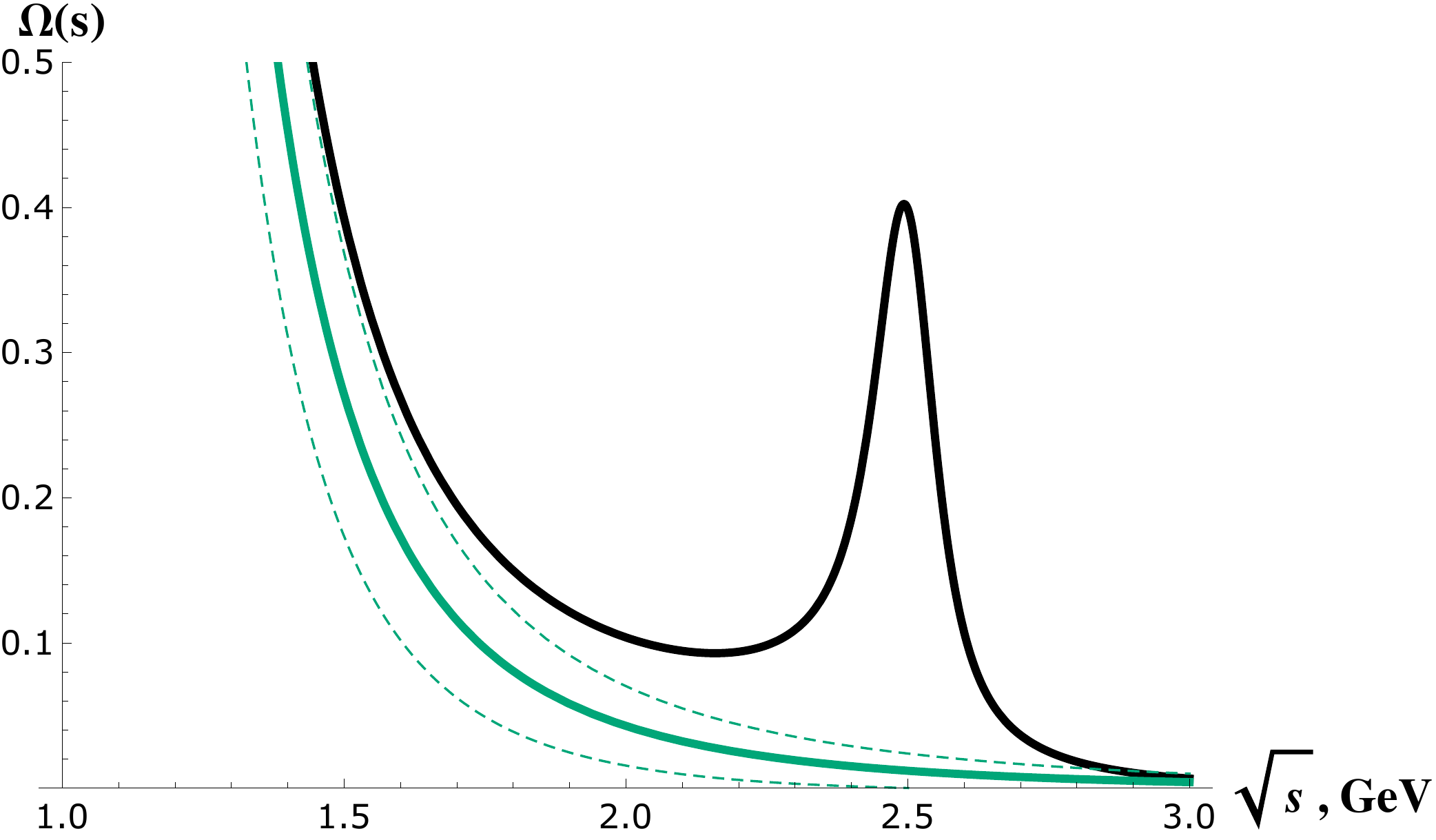}}
\ec
\caption{\label{fig:farRes} Omn\`es factors for the model in Fig.~\ref{fig:phaseModel1}: corresponding to experimental data (green), actual (black). Dashed lines represent naive corrections ($1\pm s/\Lambda_\text{diff}^2$) according to~(\ref{eq:errorEstimate}), with $\Lambda_\text{diff}=2.5$~GeV.}
\end{figure}

The first model (Fig.~\ref{fig:phaseModel1}) assumes only one relatively narrow and far (occurring at $2.5$~GeV) resonance. The corresponding Omn\`es factor, obtained using the formula (\ref{ff-solution-1ch}), is depicted in Fig.~\ref{fig:farRes}. We see that the effect of the phase change is significant only in the vicinity of the resonance. Moreover, the leading order correction~(\ref{eq:errorEstimate}) almost perfectly reproduces the result up to $1.6-1.7$~GeV.

\begin{figure}[h]
\bc
\subfloat[\label{fig:closeRes} $\Lambda_\text{diff}=1.7$~GeV.]{\includegraphics[width=0.45\textwidth]{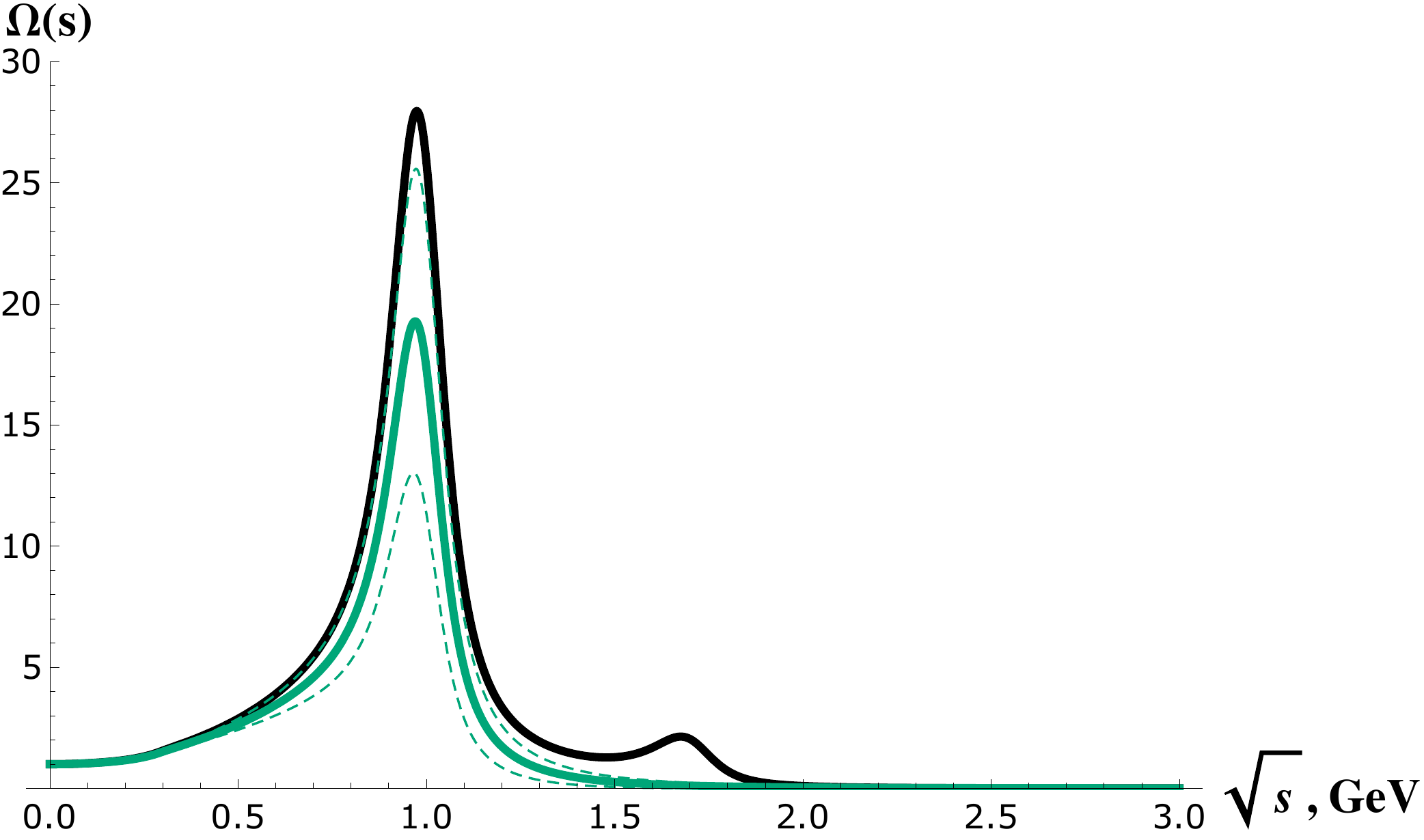}}
\hfill
\subfloat[\label{fig:2overRes} Effective $\tilde \Lambda_\text{diff}=1.3$~GeV.]{\includegraphics[width=0.45\textwidth]{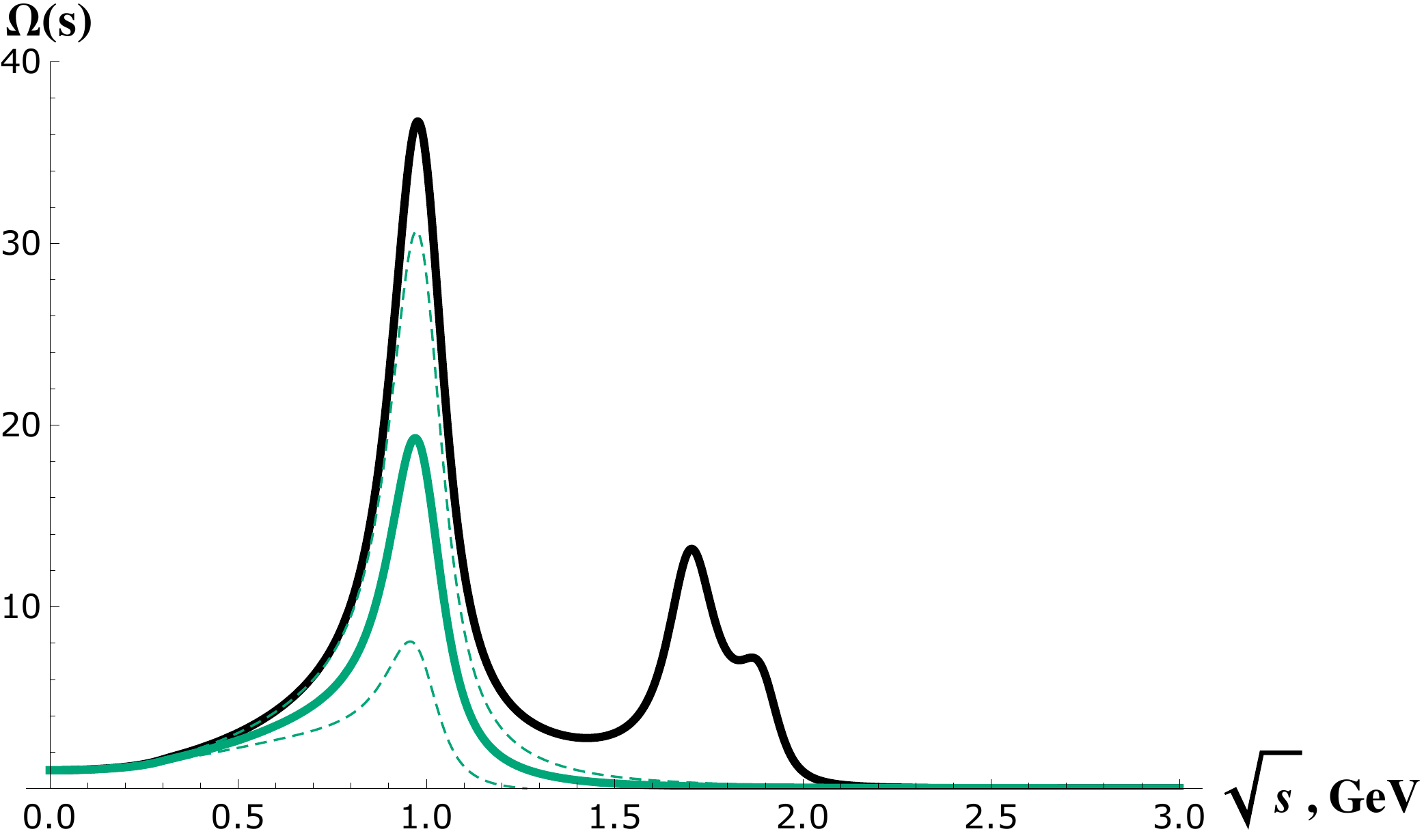}}
\ec
\caption{\label{fig:OmnesModels23} Omn\`es factors for model in Fig.~\ref{fig:phaseModel2} and~\ref{fig:phaseModel3}: corresponding to experimental data (green), actual (black). Dashed lines correspond to naively corrected results ($1\pm s/\Lambda_\text{diff}^2$) according to~(\ref{eq:errorEstimate}).}
\end{figure}

As the resonance moves closer ($1.7$~GeV), the model in Fig.~\ref{fig:phaseModel2}, it affects appreciably the form factor (see Fig.~\ref{fig:closeRes}) already around $1$~GeV and the correction~(\ref{eq:errorEstimate}) accounts for the difference at most up to $1.1-1.2$~GeV. Yet different behavior is captured by the model with two overlapping (at $1.7$ and $1.9$~GeV) resonances (Fig.~\ref{fig:phaseModel3}). The Omn\`es factor, as it is seen in Fig.~\ref{fig:2overRes},
is affected even more. The correction~(\ref{eq:errorEstimate}) should take into account both resonances now, leading to the effective $\tilde \Lambda_\text{diff}=1.6$~Gev.

What these three models show is that the Omn\`es factor obtained with incomplete data definitely cannot be extended above the corresponding cutoff $\Lambda_\text{diff}$. At the same time its low energy behavior is not affected much by the UV behavior of the scattering phase.

\bibliographystyle{utphys}
\bibliography{hadronicScalar}

\end{document}